\documentclass[a4paper,11pt]{article}
\usepackage{jheppub} % for details on the use of the package, please see the JINST-author-manual
\usepackage{lineno}
\usepackage{tikz-feynhand}
\usepackage{tikz-feynman}
\usepackage{nicematrix}
\usepackage{diagbox}
\usepackage{subcaption}
\usepackage{float}
\usepackage{physics}
\usepackage{tabularx}
\usepackage{array}
\title{\boldmath The OPE Approach to Renormalization: Operator Mixing}

\author[a]{Jinpeng Zhang}
\author[a]{, Qingjun Jin}
\affiliation[a]{Graduate School of China Academy of Engineering Physics, No. 10 Xibeiwang East Road, Haidian District, Beijing, 100193, China}
\emailAdd{zhangjinpeng23@gscaep.ac.cn}
\emailAdd{qjin@gscaep.ac.cn}

\abstract{Operator mixing is a generic and unavoidable feature in realistic quantum field theories, including QCD and effective field theories. Yet existing methods for computing anomalous dimensions in the presence of mixing remain limited by intricate sub-divergence subtractions and escalating computational complexity. In this work, we extend the recently developed OPE-based renormalization algorithm to systematically handle operator mixing. The key innovation is a recursive framework built on a strategically chosen basis of lower-dimensional symmetric traceless tensor operators as ``soft'' operators, which uniquely and fully determines the $Z$-factors of higher-dimensional ``hard'' operators from their lower-dimensional counterparts. This approach entirely eliminates the need for sub-divergence subtraction that burdens traditional methods such as the $R^{*}$ operation, thereby achieving a substantial reduction in computational complexity. As the first application of this framework, in the $\phi^{4}$ model, we have obtained the anomalous dimensions for scalar operators with $\Delta\leq 3$ to loop order five and for operators with $\Delta=4,5$ to loop order four. In the $\phi^{3}$ model, we obtain the two-loop anomalous dimensions for operators with $\Delta\leq 10$. }%These calculations, verified through self-consistency checks, establish the OPE-based algorithm as a powerful and broadly applicable tool for multi-loop renormalization in the presence of operator mixing, with a clear path toward extension to more complex theories such as QCD. }

\begin{document}
\maketitle
\flushbottom

%%%%%%%%%%%%%%%%%%%%%%%%%%%%%%%%%%%%%%%%%%%
\section{Introduction}
%%%%%%%%%%%%%%%%%%%%%%%%%%%%%%%%%%%%%%%%%%%

Composite operators are fundamental to the predictive power of QFT. In theories like QCD, physical observables are necessarily gauge-invariant combinations of the elementary quark and gluon fields. For instance, the electromagnetic structure of the proton is probed via the quark-antiquark current operator $\bar{q}\gamma^{\mu}q$, whose matrix elements define electromagnetic form factors~\cite{Peskin:1995ev,Punjabi:2015bba}. Similarly, the operator product expansion (OPE) framework uses a basis of composite operators to systematically analyze deep inelastic scattering processes, factorizing the physics into short-distance (UV) and long-distance (IR) contributions~\cite{Ellis:1978ty,Libby:1978qf,Smirnov:2002pj,Muta:2010xua}. 

The foundational role of composite operators extends across diverse theoretical landscapes. In conformal field theory (CFT), composite operators are the primary objects of study. Their correlation functions and scaling dimensions contain all physical information about the theory~\cite{DiFrancesco:1997nk,Andrianopoli:1999vr,Antipin:2021akb,Bajc:2022wws}. The AdS/CFT correspondence establishes a profound link where a composite operator in a CFT on the boundary of Anti-de Sitter (AdS) space is dual to a specific field mode in the AdS bulk~\cite{Witten:1998qj,Gubser:1998bc,Maldacena:1997re,Maldacena:2003nj}. This duality translates complex, non-perturbative strong-coupling problems in QFT into more tractable gravitational calculations~\cite{Aharony:1999ti,Harlow:2011ke,Guica:2016pid,Castro:2017hpx}. In statistical field theory, composite operators define order parameters and response functions~\cite{Marino:2017ckg}. A key example is the energy-density operator in the Ising model, which dictates the singular behavior of the specific heat near a phase transition~\cite{Cardy_1996,Caselle:1997hs}. In condensed matter physics, composite operators are essential for describing collective excitations and emergent phenomena arising from strong electron correlations, which are central to understanding phenomena like high-temperature superconductivity~\cite{Altland:2006si,Avella_2013,Di_Ciolo_2018,Irkhin_2019}. 

A defining characteristic of a composite operator is its anomalous dimension $\gamma$. This quantum correction to the operator's classical scaling dimension arises from quantum fluctuations and governs its evolution under the Renormalization Group (RG). The total scaling dimension, or conformal dimension in CFTs is $\Delta=\Delta_{0}+\gamma$, where $\Delta_{0}$ is the classical dimension. This quantity determines the power-law decay of two-point correlation functions ($\langle \mathcal{O}(x)\mathcal{O}(0) \rangle \sim 1/|x|^{2\Delta}$), and via the AdS/CFT dictionary corresponds to the mass of the dual particle in the AdS bulk \citep{Peskin:1995ev,Aharony:1999ti}. Precise calculation of anomalous dimensions is therefore a critical task across all these fields, which relies on diverse methods tailored to theoretical frameworks and coupling strengths. Non-perturbative methods for strong coupling encompass a diverse set of tools including conformal bootstrap~\cite{Poland:2018epd,Chester:2019wfx}, large-$N$ expansion~\cite{Moshe:2003xn,Lucini:2012gg}, AdS/CFT~\cite{Gubser:1998bc,Aharony:1999ti,Sakai:2004cn}, integrability~\cite{Zamolodchikov:1978xm,Gromov:2013pga,Ekhammar:2024rfj}, the lattice field theory~\cite{Wilson:1970ag,Wilson:1974sk,Gattringer:2010zz}, the functional renormalization group (FRG)~\cite{Polonyi:2001se,Dupuis:2020fhh}, and the fuzzy sphere regularization~\cite{Madore:1991bw,Zhu:2022gjc,Hofmann:2023llr,Voinea:2025iun}. 

The precision demanded by modern experiments, such as measurements of the muon's anomalous magnetic moment ($g-2$), also necessitates perturbative calculations of anomalous dimensions far beyond leading one-loop approximations~\cite{Toth:2022lsa,Davier:2023fpl}. This has spurred the development of highly sophisticated computational techniques, among them a central tool is the $R^{*}$ operation~\cite{vladimirovMethodCalculatingRenormalizationgroup1980,Chetyrkin:1981qh,Chetyrkin:1984xa,Herzog:2017bjx}, a powerful and systematic procedure for subtracting both ultraviolet (UV) and infrared (IR) divergences. It has been applied to obtain two-loop anomalous dimensions of the CP-violating Weinberg operator in full QCD, with extensions to three loops in the pure Yang-Mills limit~\cite{Brod:2019rzc,deVries:2019nsu}, and to derive anomalous dimensions of scalar and tensor operators with $\Delta \leq 6$ in the $\phi^4$ model~\cite{Henriksson:2025hwi}. Anomalous dimensions of operators in gauge theories can also be extracted from their form factors after subtraction of IR divergences~\cite{Jin:2022qjc,Blumlein:2023uuq,Jin:2023cce,Jin:2023fbz}. Although the $R^{*}$ operation has proven highly successful in multi-loop computations, it often involves significant complexity due to the intricate subtraction of sub-divergences. Furthermore, its efficiency tends to diminish for higher-dimensional operators, as the structure of sub-divergences becomes increasingly convoluted as the number of external legs in correlation functions increases. These limitations motivate the search for alternative approaches. 

Recently, a novel method for computing anomalous dimensions was developed based on the ideas of OPE and large momentum expansion~\cite{Huang:2024hsn}. This approach provides a systematic way to determine renormalization $Z$-factors by analyzing the ultraviolet finiteness of Wilson coefficients in dimensional regularization schemes. While this method has achieved record-breaking loop orders for operators like $\phi^Q$ in the $\phi^3$ model~\cite{Huang:2025rdy}, its application to the more challenging case of operator mixing has remained largely unexplored. Operator mixing is not an exceptional case but a generic feature of renormalization: any set of operators sharing the same mass dimension and quantum numbers will in general mix, and the complexity of this mixing grows with the dimension and spin of the operators~\cite{Bukhvostov:1985rn, Balitsky:1987bk, Braun:2003rp}. In effective field theory, neglecting this mixing will lead to significant errors in parameter fits. Despite its ubiquity and importance, the OPE-based method has so far been limited to the case without operator mixing.

In this paper, we fill this critical gap by extending the OPE-based renormalization algorithm to systematically treat operator mixing. The central insight is that the renormalization of all scalar operators can be organized into a closed recursive hierarchy that eliminates sub-divergence subtraction entirely. Specifically, for any scalar ``hard'' operator $\Omega_{I}$, we construct a basis of ``soft'' operators $O_{\alpha}$ from lower-dimensional symmetric traceless tensor operators. We prove that this basis yields a tree-level OPE coefficient matrix with full row rank, guaranteeing that the $Z$-factors of $\Omega_{I}$ are uniquely and fully determined. Furthermore, the renormalization of these intermediate tensor operators reduces systematically to that of even simpler scalar operators---either because they are descendants or because they can be combined with total derivatives. This establishes a recursive algorithm that computes $Z$-factors iteratively from lower to higher dimensional scalar operators, with each recursion step relying only on $Z$-factors already determined in previous steps. The entire framework requires no subtraction of UV or IR sub-divergences. 

The tadpole method provides another way to avoid explicit diagram-by-diagram subtraction of sub-divergences~\cite{Chetyrkin:1997fm}. With a suitable choice of kinematics, operator-insertion diagrams are reduced to tadpole integrals, and the ultraviolet renormalization is organized using the lower-order counterterm structure. The tadpole method and the OPE approach thus avoid explicit diagram-by-diagram subtraction of sub-divergences, but organize the ultraviolet information differently: the former extracts the relevant coefficients from vacuum integrals, whereas the latter determines them from the short-distance expansion of local operator products. 

As the first application of this framework, we report the five-loop anomalous dimensions for operators with $\Delta\leq 3$ and the four-loop anomalous dimensions for operators with $\Delta=4,5$ in the $\phi^{4}$ model and the two-loop anomalous dimensions for operators with $\Delta\leq 10$ in the $\phi^{3}$ model. These results not only demonstrate the efficiency of the OPE-based method in handling the full complexity of operator mixing, but also pave the way for extending this approach to more realistic theories, such as QCD and the Standard Model. 

The remainder of this paper is organized as follows. Section \ref{the ope approach} reviews the OPE-based renormalization framework and introduces the concepts of ``hard'' and ``soft'' operators. Section \ref{selecting O_alpha} establishes the criteria for selecting the ``soft'' operator basis and proves the full-rank property of the tree-level OPE coefficient matrix. Section \ref{demonstration} demonstrates the method through concrete examples in the $\phi^{3}$ model, including operators up to $\Delta=10$. Section \ref{re-phi3phi4} presents the five-loop results in the $\phi^{4}$ model. Section~\ref{checks} collects the comparisons with the literature and the internal consistency checks that validate the algorithm. Section \ref{conclusion} concludes with a discussion of future directions.

%%%%%%%%%%%%%%%%%%%%%%%%%%%%%%%%%%%%%%%%%%%
\section{The OPE approach to renormalization}\label{the ope approach}
%%%%%%%%%%%%%%%%%%%%%%%%%%%%%%%%%%%%%%%%%%%
The operator product expansion~\cite{Wilson:1969zs} describes the behavior of the products of local operators at short or light-like distances, and has found remarkably successful applications in QFT~\cite{Shifman:1978bx, Shifman:1978by,Chetyrkin:1985kn,
Pich:1999hc,Pich:2016bdg,
Boito:2020xli,Ayala:2022cxo,
Bruser:2024zyg} and CFT~\cite{Luscher:1975js,Dolan:2000ut,
Dolan:2003hv,Hollands:2006ag,
Holland:2014ifa,Gillioz:2019iye,
Hollands:2023txn}. As the separation $x\rightarrow 0$, the OPE for two operators takes the schematic form 
\begin{equation}\label{simple-ope}
    \begin{aligned}
        O_{I}(x)O_{J}(0)\sim\sum_{\alpha}{\tilde{C}_{IJ}}^{\alpha}(x)O_{\alpha}(0)\ , 
    \end{aligned}
\end{equation}
where $O_{\alpha}(0)$ denote generic local operators with the same symmetry as the left-hand side (LHS) of \eqref{simple-ope}, and ${\tilde{C}_{IJ}}^{\alpha}(x)$ refers to the OPE coefficient corresponding to $O_{\alpha}(0)$. The OPE coefficients are proven finite \cite{Chetyrkin:1982zq,Chetyrkin:1988zz} and can be reduced to propagator-type integrals \cite{Tkachov:1983st}. Systematic algorithms for computing these coefficients perturbatively in the MS scheme were proposed in \cite{Chetyrkin:1982zq, Gorishnii:1983su, Gorishnii:1986gn,LlewellynSmith:1987jx}. 

In momentum space, the LHS of \eqref{simple-ope} involves the product of two operators with large momentum $p$, referred to as hard operators: 
\begin{equation}
    \begin{aligned}
        O_{I}(p)O_{J}(k-p)\sim \sum_{\alpha} {C_{IJ}}^{\alpha}(p) O_{\alpha}(k)\ . 
    \end{aligned}
\end{equation}
On the right-hand side (RHS), all dependence on the large momentum $p$ is encapsulated within the OPE coefficients, while the resulting operators possess small momentum $k$ and are therefore designated as soft operators. 

It has long been established that the renormalization flow of OPE coefficients is governed by the anomalous dimensions of both hard and soft operators. For example, in the $\phi^{4}$ model, the OPE of two fundamental fields reads 
\begin{equation}
    \begin{aligned}
        \phi(p)\phi(k-p)\sim C_{\phi^{2}}(p)\phi^{2}_{R}(k)+\cdots\ , 
    \end{aligned}
\end{equation}
and renormalization group flow equation (RGE) of the OPE coefficient $C_{\phi^{2}}$ is given by~\cite{Collins:1984xc}: 
\begin{equation}
    \begin{aligned}
        \frac{\partial C_{\phi^{2}}}{\partial \ln\mu}=(\gamma_{\phi^{2}}+2\gamma_{\phi})C_{\phi^{2}}\ . 
    \end{aligned}
\end{equation}
Given that $\gamma_{\phi}$ (the anomalous dimension of the fundamental field) is already known, $\gamma_{\phi^{2}}$ (the anomalous dimension of composite operator $\phi^{2}$) can be derived from $C_{\phi^{2}}$. 

Nevertheless, this method is less efficient for evaluating $\gamma_{\phi^{2}}$, as it requires extracting $C_{\phi^{2}}$ from the four-point correlation function $\langle \phi\phi\phi\phi\rangle$. In contrast, the standard and more efficient approach determines $\gamma_{\phi^{2}}$ directly from the following three-point correlation function: 
\begin{equation}
    \begin{aligned}
        \mathcal{G}(p;k)\equiv \Bigl\langle \phi^{2}_{R}(p)\phi(-k-p)\phi(k)\Bigr\rangle\ . 
    \end{aligned}
\end{equation}

By applying OPE we can further simplify this standard approach. Specifically, we consider the OPE of $\phi^{2}_{R}$ and $\phi$: 
\begin{equation}
    \begin{aligned}
        \phi^{2}_{R}(p)\phi(-k-p)\sim C(p)\phi(k)+\cdots\ . 
    \end{aligned}\label{phi2OPE}
\end{equation}
In this context, the OPE coefficient $C(p)$ can be directly obtained from the three-point correlation function $\mathcal{G}(p;k)$. Furthermore, $C(p)$ enables the determination of $\gamma_{\phi^{2}}$ via the following RGE: 
\begin{equation}
    \begin{aligned}
        \frac{\partial C(p)}{\partial \ln\mu}=-\gamma_{\phi^{2}}C(p)\ . 
    \end{aligned}
\end{equation}
This OPE-based strategy, which we will elaborate on in the subsequent part of this section, can be generalized to investigate the renormalization of arbitrary operators in general quantum field theories.

%%%%%%%%%%%%%%%%%%%%%%%%%%%%%%%%%%%%%%%%%%%
\subsection{OPE in the situation of operator mixing}\label{ope-o-phi}
%%%%%%%%%%%%%%%%%%%%%%%%%%%%%%%%%%%%%%%%%%%
Let us consider the renormalization of a set of mixed scalar operators $\Omega_{I}^{R}$ with classical scaling dimension $\Delta_0$ in $\phi^{4}$ model. %The supercript $R$ indicates that the operators are renormalized. 
Their anomalous dimensions can be extracted from the correlation functions of $\Omega_{I}^{R}$ and several fundamental fields: 
\begin{equation}
    \begin{aligned}
        \mathcal{G}_{I}(p;k_{1},\cdots, k_{n})\equiv \Bigl\langle \Omega_{I}^{R}(p)\phi(k-p)\phi(k_{1})\cdots\phi(k_{n})\Bigr\rangle\ , 
    \end{aligned}
\end{equation}
where $n$ satisfies $n\leq \Delta_0$, and $k=-\sum_{i=1}^{n}k_{i}$. 

The degree of difficulty to evaluate the contributing Feynman integrals depends on the number of scales in the integrals. To reduce the number of scales, we assume that in the Euclidean sense, $|p|\gg |k_{i}|$, where $|p|=\sqrt{p^{2}_{0}+\mathbf{p}^{2}}$~\cite{Smirnov:1990rz,Smirnov:1994tg,Smirnov:2002pj}, so that the hard and soft dynamics factorize via the following OPE: 
\begin{equation}\label{ope-op}
    \begin{aligned}
        \Omega_{I}^{R}(p)\phi(k-p)\sim \sum_{\alpha}{C_{I}}^{\alpha}(p)O^{R}_{\alpha}(k)+\cdots\ , 
    \end{aligned}
\end{equation}
in which $O^{R}_{\alpha}$ is another set of (preferably simpler) mixed operators. Then $\Omega_{I}^{R}$ are hard operators, and $O^{R}_{\alpha}$ are soft operators, since they carry hard momentum $p$ and soft momentum $k$, respectively. Since both $\Omega_{I}^{R}$ and $O^{R}_{\alpha}$ are renormalized operators, the OPE coefficients ${C_{I}}^{\alpha}$ must be UV finite, which imposes constraints on the $Z$-factors. 

To be more concrete, the $Z$-factors of $\Omega_{I}^{R}$ and $O^{R}_{\alpha}$ are given by 
\begin{equation}
    \begin{aligned}
        \Omega_{I}^{R}={Z_{I}}^{J}\Omega_{J}^{0},\ O_{\alpha}^{R}={Z_{\alpha}}^{\beta}O_{\beta}^{0}\ , 
    \end{aligned}
\end{equation}
in which $\Omega_{J}^{0}$ and $O_{\beta}^{0}$ are bare operators consisting of bare field and coupling, $\phi_{0}$ and $g_{0}$. For example, the $\Delta=3$ bare operators can be chosen as 
\begin{equation}\label{Z-define}
    \begin{aligned}
        \Omega_{I}^{0}=(\partial^{2}\phi_{0},\ \frac{g_{0}}{3!}\phi_{0}^{3})\ . 
    \end{aligned}
\end{equation}

The product of bare operators also permit a OPE relation: 
\begin{equation}\label{ope-op-bare}
    \begin{aligned}
        \Omega_{I}^{0}(p)\phi_{0}(k-p)\sim \sum_{\alpha}{C_{0I}}^{\alpha}(p)O^{0}_{\alpha}(k)+\cdots\ . 
    \end{aligned}
\end{equation}
Using \eqref{Z-define} and $\phi_{0}=Z_{\phi}^{\frac{1}{2}}\phi$, and comparing \eqref{ope-op} and \eqref{ope-op-bare}, the relation between the bare and renormalized OPE coefficients reads\footnote{The renormalization flow of ${C_{I}}^{\alpha}$ is given by
\begin{equation*}
    \begin{aligned}
        \frac{\partial {C_{I}}^{\alpha}(p)}{\partial\ln\mu}
        =-\gamma_{\phi}{C_{I}}^{\alpha}(p)-{\gamma_{I}}^{J}{C_{J}}^{\alpha}(p)+{C_{I}}^{\beta}(p){\gamma_{\beta}}^{\alpha}\ , 
    \end{aligned}
\end{equation*}
in which ${\gamma_{I}}^{J}$ and ${\gamma_{\beta}}^{\alpha}$ are the anomalous dimension matrices for $\Omega_{I}$ and $O_{\alpha}$, respectively: 
\begin{equation*}
    \begin{aligned}
        {\gamma_{I}}^{J}=-\frac{\partial {Z_{I}}^{K}}{\partial\ln\mu}{(Z^{-1})_{K}}^{J},\ {\gamma_{\beta}}^{\alpha}=-\frac{\partial {Z_{\beta}}^{\rho}}{\partial\ln\mu}{(Z^{-1})_{\rho}}^{\alpha}\ . 
    \end{aligned}
\end{equation*}} 
\begin{equation}\label{coe-relation}
    \begin{aligned}
        {C_{I}}^{\alpha}(p)=Z_{\phi}^{-\frac{1}{2}}{Z_{I}}^{J}{C_{0J}}^{\beta}(p){(Z^{-1})_{\beta}}^{\alpha}\ . 
    \end{aligned}
\end{equation}
Suppose $Z_{\phi}$ and ${Z_{\beta}}^{\alpha}$ are already known, the UV finiteness of ${C_{I}}^{\alpha}(p)$ impose constraints on ${Z_{I}}^{J}$. Different choice of $O^{R}_{\alpha}$ yields distinct constraints, and the $Z$-factors can be fully determined by considering a sufficient set of $O^{R}_{\alpha}$. 

In this work we will focus on the case that the hard operators $\Omega_{I}^{R}$ are Lorentz scalars, then the soft operators $O_{\alpha}$ are necessarily traceless symmetric tensors. This conclusion can be established through the following reasoning. First, decompose all soft operators into irreducible representations of the Lorentz group $SO(1,D-1)$. By definition, these irreducible representations correspond exclusively to traceless tensors. Since the operators on the LHS of \eqref{ope-op} and \eqref{ope-op-bare} are Lorentz scalars, every Lorentz index in each tensor operator which appears on the RHS must be contracted with the associated OPE coefficient. Within the OPE framework, such tensor indices can only be carried by hard momenta (denoted by $p^{\mu}$) or the Minkowski metric (denoted by $\eta^{\mu\nu}$). Notably, terms involving $\eta^{\mu\nu}$ can always be discarded, because the contraction of $\eta^{\mu\nu}$ with a traceless tensor operator vanishes identically. With metric terms eliminated, the OPE coefficients for rank-$J$ tensor operators are proportional solely to expressions of the form $(p^{2})^rp^{\mu_{1}}\cdots p^{\mu_{J}}$, which is fully symmetric across all its Lorentz indices. When contracted with these symmetric OPE coefficients, only the fully symmetric tensor operators remain non-vanishing. Any tensors belonging to other irreducible representations of $SO(1,D-1)$ (e.g., antisymmetric or mixed-symmetry tensors) are canceled out entirely. Collectively, this reasoning confirms that the soft operators on the OPE's RHS must be traceless symmetric tensors. 

As a side remark, the information regarding $Z$-factors can also be obtained by examining the OPE of two composite operators. For example, the $Z$-factor of $\phi^3$ can be obtained by analyzing 
\begin{equation}
    \begin{aligned}
        \phi^3(p)\phi^2(k-p)\sim C(p)\phi(k)+\cdots\ .
    \end{aligned}
\end{equation}
However, the leading-order contribution of $C(p)$ already involves loop integrals. To compute the $L$-loop correction, one must evaluate $(L+1)$-loop integrals. This significantly increases the computational difficulty compared to the OPE involving fundamental fields. Therefore, we will not consider the OPE of two composite operators in this context.

%%%%%%%%%%%%%%%%%%%%%%%%%%%%%%%%%%%%%%%%%%%
\subsection{Projection operators}
%%%%%%%%%%%%%%%%%%%%%%%%%%%%%%%%%%%%%%%%%%%

OPE coefficients can be extracted by applying projection operators to correlation functions of $\Omega_I$. These projectors filter out contributions of all other soft operators, leaving only the term proportional to the target OPE coefficients~\cite{Gorishnii:1983su,Gorishnii:1986gn,Smirnov:2002pj}. 

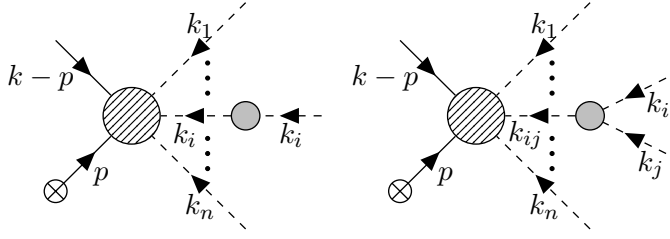
\begin{figure}
	\centering
		\subfloat{
		\begin{tikzpicture}
		\begin{feynhand}
			\vertex [NEblob] (a0) at (-2,0) {};
			\vertex [crossdot] (a1) at (-3,-1) {};
			\propag[fer] (a1) to (a0);
			\node at (-2.4,-0.8) {$p$};
			\setlength{\feynhanddotsize}{0.5mm}
			\vertex (a2) at (-3,1);
			\propag[fer] (a2) to (a0);
			\node at (-3.2,0.5) {$k-p$};
			\vertex (b1) at (0.5,0);
			\vertex[grayblob, style={scale=0.5}] (b2) at (-0.5,0) {};
			\vertex (b3) at (-0.5,1.5);
			\vertex (b4) at (-0.5,-1.5);
			\propag[chasca] (b1) to (b2);
			\propag[chasca] (b2) to (a0);
			\propag[chasca] (b3) to (a0);
			\propag[chasca] (b4) to (a0);
			\vertex [dot] (c1) at (-1,0.3) {};
			\vertex [dot] (c2) at (-1,0.5) {};
			\vertex [dot] (c3) at (-1,0.7) {};
			\vertex [dot] (c4) at (-1,-0.3) {};
			\vertex [dot] (c5) at (-1,-0.5) {};
			\vertex [dot] (c6) at (-1,-0.7) {};
			\node at (-1.1,1.2) {$k_{1}$};
			\node at (-1.1,-1.2) {$k_{n}$};
			\node at (0.1,-0.25) {$k_{i}$};
			\node at (-1.3,-0.25) {$k_{i}$};
		\end{feynhand}
		\end{tikzpicture}
		}
		\subfloat{
			\begin{tikzpicture}
			\begin{feynhand}
				\vertex [NEblob] (a0) at (-2,0) {};
				\vertex [crossdot] (a1) at (-3,-1) {};
				\propag[fer] (a1) to (a0);
				\node at (-2.4,-0.8) {$p$};
				\setlength{\feynhanddotsize}{0.5mm}
				\vertex (a2) at (-3,1);
				\propag[fer] (a2) to (a0);
				\node at (-3.2,0.5) {$k-p$};
				\vertex (b1) at (0.5,0.5);
				\vertex (b5) at (0.5,-0.5);
				\vertex[grayblob, style={scale=0.5}] (b2) at (-0.5,0) {};
				\vertex (b3) at (-0.5,1.5);
				\vertex (b4) at (-0.5,-1.5);
				\propag[chasca] (b1) to (b2);
				\propag[chasca] (b5) to (b2);
				\propag[chasca] (b2) to (a0);
				\propag[chasca] (b3) to (a0);
				\propag[chasca] (b4) to (a0);
				\vertex [dot] (c1) at (-1,0.3) {};
				\vertex [dot] (c2) at (-1,0.5) {};
				\vertex [dot] (c3) at (-1,0.7) {};
				\vertex [dot] (c4) at (-1,-0.3) {};
				\vertex [dot] (c5) at (-1,-0.5) {};
				\vertex [dot] (c6) at (-1,-0.7) {};
				\node at (-1.1,1.2) {$k_{1}$};
				\node at (-1.1,-1.2) {$k_{n}$};
				\node at (0.4,0.15) {$k_{i}$};
				\node at (0.3,-0.65) {$k_{j}$};
				\node at (-1.35,-0.25) {$k_{ij}$};
			\end{feynhand}
		    \end{tikzpicture}
		}
\caption{The diagram with $\frac{1}{k_i^2}$ or $\frac{1}{(k_i+k_j)^2}$ type propagators, where $k_{ij}=k_i+k_j$. The dashed lines correspond to external legs carrying soft momenta, and we will adhere to this convention in the rest of the paper. }
\label{fig:excluded}
\end{figure}

Let us first introduce the concept of hard correlation function. The hard correlation function, denoted by
\begin{equation}
    \begin{aligned}
        \mathcal{H}_I(p;k_1,\cdots, k_n)\equiv \Bigl\langle \Omega_{I}^{R}(p)\phi(k-p)\phi(k_1)\cdots \phi(k_n)\Bigr\rangle_{h}\ , 
    \end{aligned}
\end{equation}
receives contribution exclusively from the Feynman diagrams where all internal propagators carry hard momenta. Consequently, diagrams containing propagators of the form $\frac{1}{k_i^2}$ or $\frac{1}{(k_i+k_j)^2}$, as illustrated in Figure \ref{fig:excluded}, are excluded from consideration. All contributing Feynman diagrams consistently exhibit a chain-like structure, which will be analyzed in detail in Section \ref{section:1PR}. 

The bare OPE coefficient corresponding to a length-$n$ soft operator $O_{\alpha}^0$ can be obtained by acting a projection operator $\mathcal{P}_{0n}^{\alpha}$ on the hard bare correlation function $\mathcal{H}_{0I}$~\cite{Gorishnii:1983su,Gorishnii:1986gn}:
\begin{equation}
    \begin{aligned}
        {\mathcal{C}_{0I}}^{\alpha}(p)=\mathcal{P}_{0n}^{\alpha}\mathcal{H}_{0I}(p;k_1,\cdots, k_n)\Bigr|_{k_i=0}\ . 
    \end{aligned}
\end{equation}
The explicit form of the projection operator can be determined through the following tree-level correlation function: 
\begin{equation}
    \begin{aligned}
        F_{\alpha}(k_1,\cdots, k_n)=\Bigl\langle O_{\alpha}(k)\phi(k_1)\cdots \phi(k_n)\Bigr\rangle^{\text{tree}}\ . 
    \end{aligned}
\end{equation}
This quantity is nothing more than the Feynman rules of the $O_{\alpha}\phi^n$ vertex in momentum space, and we will call it the minimal form factor for brevity. $\mathcal{P}_n^{\alpha}$ is a differential operator composed of $\hat{\partial}_{i\mu}\equiv \frac{\partial}{\partial k_{i}^{\mu}}$, and satisfies
\begin{equation}
    \begin{aligned}
        \mathcal{P}_{n}^{\alpha}F_{\beta}(k_{1},\cdots ,k_{n})=\delta^{\alpha}_{\beta}\ . 
    \end{aligned}
\end{equation}
It can be interpreted as the dual of $F_{\beta}$. 

By substituting $k_{i}\rightarrow \hat{\partial}_{i}$, each minimal form factor is mapped to a differential operator: 
\begin{equation}
    \begin{aligned}
        F_{\alpha}(k_{i})\rightarrow F_{\alpha}(\hat{\partial}_{i})\ . 
    \end{aligned}
\end{equation}
Since $F_{\alpha}(k_{i})$ form a basis for $k_{i}$-polynomials, $F_{\alpha}(\hat{\partial}_{i})$ also form a basis for $\hat{\partial}_{i}$-polynomials. As a result, the projection operators can always be expanded in terms of $F_{\alpha}(\hat{\partial}_{i})$: 
\begin{equation}
    \begin{aligned}
        \mathcal{P}_{n}^{\alpha}=\sum_{\beta}M^{\alpha\beta}F_{\beta}(\hat{\partial}_{i})\ , 
    \end{aligned}
\end{equation}
where $M^{\alpha\beta}$ is the inverse of
\begin{equation}
    \begin{aligned}
        M_{\alpha\beta}=F_{\alpha}(\hat{\partial}_{i})F_{\beta}(k_i)\ , 
    \end{aligned}
\end{equation}
which can be interpreted as the inner product of $F_{\alpha}$ and $F_{\beta}$. 

As discussed in Section \ref{ope-o-phi}, traceless symmetric tensor operator emerge as soft operators in the OPE of two Lorentz scalar operators. The projection operator associated with a rank-$J$ traceless symmetric tensor operator $O_{\alpha}^{\mu_{1}\cdots\mu_{J}}$ can be constructed using the same methodology applied to the scalar operator case (i.e. $J=0$). The corresponding minimal form factor will be denoted by $F_{\alpha}^{\mu_{1}\cdots\mu_{J}}(k_{i})$, and the projection operator $\mathcal{P}^{\alpha}_{n;\mu_{1}\cdots\mu_{J}}(\hat{\partial}_{i})$ satisfies the relation: 
\begin{equation}\label{tensor-projection}
    \begin{aligned}
        \mathcal{P}^{\alpha}_{n;\mu_{1}\cdots\mu_{J}}(\hat{\partial}_{i})F_{\alpha}^{\nu_{1}\cdots\nu_{J}}(k_{i})=\delta_{(\mu_{1}}^{\nu_{1}}\cdots \delta_{\mu_{J})}^{\nu_{J}}\ , 
    \end{aligned}
\end{equation}
where $\delta_{(\mu_{1}}^{\nu_{1}}\cdots \delta_{\mu_{J})}^{\nu_{J}}$ represents the symmetric traceless component of $\delta_{\mu_{1}}^{\nu_{1}}\cdots \delta_{\mu_{J}}^{\nu_{J}}$. While the symmetrization and traceless conditions are imposed on the indices $\mu_{i}$, this quantity also acts as a symmetric traceless tensor with respect to the indices $\nu_{i}$. For example, 
\begin{equation}
    \begin{aligned}
        \delta_{(\mu_{1}}^{\nu_{1}} \delta_{\mu_{2})}^{\nu_{2}}=\frac{1}{2}(\delta_{\mu_{1}}^{\nu_{1}}\delta_{\mu_{2}}^{\nu_{2}}+\delta_{\mu_{1}}^{\nu_{2}}\delta_{\mu_{2}}^{\nu_{1}})-\frac{1}{D}\eta_{\mu_{1}\mu_{2}}\eta^{\nu_{1}\nu_{2}}\ , 
    \end{aligned}
\end{equation}
where $D$ is the dimension of spacetime. By contracting both sides of \eqref{tensor-projection} with $\delta_{\nu_{1}}^{\mu_{1}}\cdots \delta_{\nu_{J}}^{\mu_{J}}$, we obtain 
\begin{equation}\label{tensor-projection-1}
    \begin{aligned}
        \mathcal{P}^{\alpha}_{n;\mu_{1}\cdots \mu_{J}}(\hat{\partial}_{i})F_{\alpha}^{\mu_{1}\cdots\mu_{J}}(k_{i})=c_{J}\ , 
    \end{aligned}
\end{equation}
in which $c_{J}$ is given by: 
\begin{equation}
    \begin{aligned}
        c_{J}=\frac{(D+2J-2)(D-1)_{J}}{(D+J-2)(J!)^{2}}\ , 
    \end{aligned}
\end{equation}
where $(D-1)_{J}$ is the Pochhammer symbol. The inner product of minimal form factors is defined as follows:
\begin{equation}\label{tensor-inner}
    \begin{aligned}
        M_{J;\alpha\beta}=F_{\alpha}^{\mu_{1}\cdots\mu_{J}}(\hat{\partial}_{i})F_{\beta;\mu_{1}\cdots\mu_{J}}(k_{i})\ . 
    \end{aligned}
\end{equation}
Combining \eqref{tensor-projection-1} and \eqref{tensor-inner} allows us to derive the expression for the tensor projection operator: 
\begin{equation}
    \begin{aligned}
        \mathcal{P}^{\alpha}_{n;\mu_{1}\cdots \mu_{J}}(\hat{\partial}_{i})=c_{J}M_{J}^{\alpha\beta}F_{\beta;\mu_{1}\cdots\mu_{J}}(\hat{\partial}_{i})\ , 
    \end{aligned}
\end{equation}
where $M_{J}^{\alpha\beta}$ is the inverse of $M_{J;\alpha\beta}$.

The integrals that contribute to OPE coefficients are two-point integrals characterized by a single external momentum scale, $p^2$. These integrals can be simplified into master integrals through tensor reduction and integration by parts (IBP)~\cite{Tkachov:1981wb,Chetyrkin:1981qh,Derkachov:1997ch,Laporta:2000dsw,Smirnov:2019qkx,Cao:2021cdt,Wu:2023upw,Guan:2024byi}. The analytic expressions for these master integrals are known through five-loop order~\cite{Georgoudis:2021onj}. At four loops, all master integrals were evaluated algebraically in Ref.~\cite{Baikov:2010hf}, the results up to transcendentality weight twelve were obtained in Ref.~\cite{Lee:2011jt}, and dedicated parametric reduction programs such as FORCER are available~\cite{Ruijl:2017cxj}. 
%The analytic expressions for these master integrals are available up to five-loop order~\cite{Georgoudis:2021onj}, for four-loop computations, see also Refs.~\cite{Baikov:2010hf,Lee:2011jt,Ruijl:2017cxj}. 

%%%%%%%%%%%%%%%%%%%%%%%%%%%%%%%%%%%%%%%%%%%
\subsection{Amputated OPE coefficients}
%%%%%%%%%%%%%%%%%%%%%%%%%%%%%%%%%%%%%%%%%%%

In practice, it is more convenient to compute the amputated correlation functions instead of the full correlation function. The amputation of the external leg $\phi(k-p)$ effectively separates a 2-point function from the full correlation function: 
\begin{equation}
    \begin{aligned}
        \mathcal{H}_{0I}(p;k_1,\cdots, k_n)=& H_{0I}(p;k_1,\cdots, k_n)G_{0}(k-p)\ ,\\
        \mathcal{H}_{I}(p;k_1,\cdots, k_n)=& H_{I}(p;k_1,\cdots, k_n)G(k-p)\ , 
    \end{aligned}
\end{equation}
where $H_{0I}$ and $H_{I}$ represent the amputated correlation functions, and $G_{0}$ and $G$ are the respective two-point correlation functions defined as: 
\begin{equation}
    \begin{aligned}
        G_{0}(k)=&\Bigl\langle \Phi_0(k)\Phi_0(-k)\Bigr\rangle=Z_{\phi}G(k)\ ,  \\
        G(k)=&\Bigl\langle \Phi(k)\Phi(-k)\Bigr\rangle\ . 
    \end{aligned}
\end{equation}
There is no need to amputate the soft external legs $\phi(k_i)$, because by construction the hard correlation function does not contain any two-point functions attached to these soft external legs. 

The amputated OPE coefficient ${C_{0I}}^{\alpha}$ is derived by acting the corresponding projection operator on the amputated hard correlation function: 
\begin{equation}
    \begin{aligned}
        {C_{0I}}^{\alpha}(p)=\mathcal{P}_{0n}^{\alpha}H_{0I}(p;k_1,\cdots, k_n)\Bigr|_{k_{i}=0}\ . 
    \end{aligned}
\end{equation}
Effectively, these amputated OPE coefficients can be regarded as the coefficients of the following amputated OPE: 
\begin{equation}\label{amputate-ope}
    \begin{aligned}
        \Omega_{I}^{R}(p)\phi(k-p)\sim &G(k-p)\sum_{\alpha}{C_{I}}^{\alpha}(p)O^{R}_{\alpha}(k)\ ,\\
        \Omega_{I}^{0}(p)\phi_{0}(k-p)\sim &G_{0}(k-p)\sum_{\alpha}{C_{0I}}^{\alpha}(p)O^{0}_{\alpha}(k)\ . 
    \end{aligned}
\end{equation}

By comparing these expressions, one can establish the relationship between the amputated OPE coefficients:
\begin{equation}\label{ope-coe-relation}
    \begin{aligned}
        {C_{I}}^{\alpha}(p)=Z_{\phi}^{\frac{1}{2}}{Z_{I}}^{J}{C_{0J}}^{\beta}(p){(Z^{-1})_{\beta}}^{\alpha}\ . 
    \end{aligned}
\end{equation}
In $Z_{\phi}$, the exponent changes from $-\frac{1}{2}$ in \eqref{coe-relation} to $\frac{1}{2}$. From this point onward, the term ``OPE coefficients'' shall always refer to these amputated OPE coefficients throughout this paper.

%%%%%%%%%%%%%%%%%%%%%%%%%%%%%%%%%%%%%%%%%%%
\subsection{The Chain structure of 1PR diagrams\label{section:1PR}}
%%%%%%%%%%%%%%%%%%%%%%%%%%%%%%%%%%%%%%%%%%%
The Feynman diagrams that contribute to amputated hard correlation functions can be classified as either one-particle irreducible (1PI) or one-particle reducible (1PR)~\cite{Gorishnii:1986gn}. A 1PR diagram can be constructed from multiple 1PI diagrams, indicating that 1PI diagrams serve as the fundamental building blocks of the correlation function.

In the conventional approach, the UV finiteness of the correlation function is ensured by the UV finiteness of the 1PI building blocks. Consequently, only 1PI diagrams require detailed investigation in practice.
In contrast, the OPE approach necessitates the simultaneous consideration of both 1PI and 1PR diagrams to ensure UV finiteness. 

\begin{figure}[htbp]
    \centering
    \subfloat{
    \begin{tikzpicture}
        \begin{feynhand}
            \vertex (a1) at (-0.5,0.5);
            \vertex (a2) at (0.5,0.5);
            \vertex (a3) at (0.5,-0.5);
            \vertex (a4) at (-0.5,-0.5);
            \propag[plain] (a1) to (a2);
            \propag[plain] (a2) to (a3);
            \propag[plain] (a3) to (a4);
            \propag[plain] (a4) to (a1);
            \vertex (b1) at (-1.0,1.0);
            \vertex (b2) at (1.0,1.0);
            \vertex (b3) at (1.0,-1.0);
            \vertex (b4) at (-1.0,-1.0);
            \propag[chasca] (b1) to (a1);
            \node at (-1.05,0.7) {$k_{1}$};
            \propag[chasca] (b2) to (a2);
            \node at (1.1,0.7) {$k_{2}$};
            \propag[fer] (b3) to (a3);
            \node at (-1.05,-0.7) {$p$};
            \propag[fer] (b4) to (a4);
            \node at (1.35,-0.7) {$k-p$};
        \end{feynhand}
    \end{tikzpicture}
    }
    \quad\quad\quad\quad\quad
    \subfloat{
    \begin{tikzpicture}
        \begin{feynhand}
            \vertex (a1) at (-0.5,0.5);
            \vertex (a2) at (0.5,0.5);
            \vertex (a3) at (0.5,-0.5);
            \vertex (a4) at (-0.5,-0.5);
            \propag[plain] (a1) to (a2);
            \propag[plain] (a2) to (a3);
            \propag[plain] (a3) to (a4);
            \propag[plain] (a4) to (a1);
            \vertex (b1) at (-1.0,1.0);
            \vertex (b2) at (1.0,1.0);
            \vertex (b3) at (1.0,-1.0);
            \vertex (b4) at (-1.0,-1.0);
            \propag[chasca] (b1) to (a1);
            \node at (-1.05,0.7) {$k_{1}$};
            \propag[fer] (b2) to (a2);
            \node at (1.35,0.7) {$k-p$};
            \propag[chasca] (b3) to (a3);
            \node at (1.1,-0.7) {$k_{2}$};
            \propag[fer] (b4) to (a4);
            \node at (-1.05,-0.7) {$p$};
        \end{feynhand}
    \end{tikzpicture}
    }
    \quad\\
    \vspace{0.5cm}
    \subfloat{
    \begin{tikzpicture}
        \begin{feynhand}
            \vertex (a1) at (0,0);
            \vertex (a2) at (1.5,0);
            \vertex (a3) at (2.5,1);
            \vertex (a4) at (2.5,-1);
            \vertex (a5) at (-0.5,-0.5);
            \vertex (a6) at (-0.5,0.5);
            \propag[plain] (a1) to (a2);
            \propag[chasca] (a3) to (a2);
            \node at (2.4,0.5) {$k_{2}$};
            \propag[fer] (a4) to (a2);
            \node at (2.6,-0.5) {$k-p$};
            \propag[plain] (a5) to (a1);
            \propag[plain] (a6) to (a1);
            \propag[plain] (a5) to (a6);
            \vertex (a7) at (-1,1);
            \vertex (a8) at (-1,-1);
            \propag[chasca] (a7) to (a6);
            \node at (-1,0.6) {$k_{1}$};
            \node at (-1,-0.6) {$p$};
            \propag[fer] (a8) to (a5);
        \end{feynhand}
    \end{tikzpicture}
    }
    \quad
    \subfloat{
    \begin{tikzpicture}
        \begin{feynhand}
            \vertex [ringblob] (a0) at (0,0) {};
            \vertex (a1) at (-1,0);
            \vertex (a2) at (1.0,0);
            \propag[plain] (a1) to (a0);
            \propag[plain] (a2) to (a0);
            \vertex (a3) at (-2,1);
            \vertex (a4) at (-2,-1);
            \vertex (a5) at (2,1);
            \vertex (a6) at (2,-1);
            \propag[chasca] (a3) to (a1);
            \node at (-1.8,0.5) {$k_{1}$};
            \propag[fer] (a4) to (a1);
            \node at (-1.8,-0.5) {$p$};
            \propag[chasca] (a5) to (a2);
            \node at (2.0,0.5) {$k_{2}$};
            \propag[fer] (a6) to (a2);
            \node at (2.1,-0.5) {$k-p$};
        \end{feynhand}
    \end{tikzpicture}
    }
    \caption{Diagrams which contribute to $C_0(g_0;p)$, diagrams on the first line correspond to the 1PI contribution and diagrams on the second line correspond to the 1PR contribution. }
    \label{fig:example-1PR}
\end{figure}
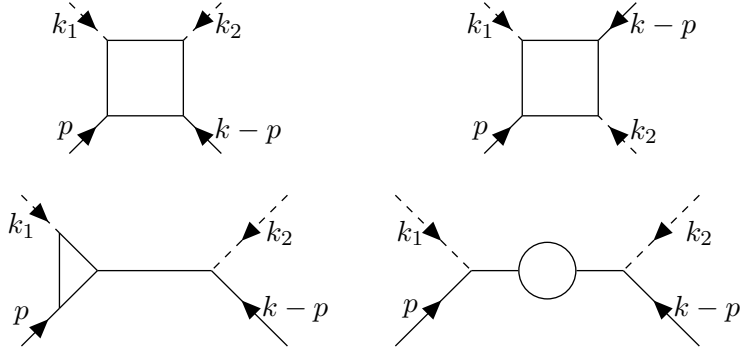

For example, we consider the following OPE in $\phi^3$ model,
\begin{equation}
\phi(p)\phi(k-p)\sim \frac{1}{2}C(p)\phi^2(k)\ . 
\end{equation}
The one-loop bare Wilson coefficient $C_0(g_0;p)$ receives contributions from one 1PI topology and two 1PR topology, as illustrated in Figure \ref{fig:example-1PR}. It can be expressed as:
\begin{equation}
C_0(g_0;p) = C_0^{\text{1PI}} + C_0^{\text{1PR}}\ , 
\end{equation}
where
\begin{equation}
\begin{aligned}
C_0^{\text{1PI}}=&\Bigl\langle \phi_0(p)\phi_0(k-p)\phi_0(k_1)\phi_0(k_2)\Bigr\rangle^{\text{1PI}}_{k_i\rightarrow 0}\ , \\
C_0^{\text{1PR}}=&\Bigl\langle \phi_0(p)\phi_0(k_1)\phi_0(-k_1-p)\Bigr\rangle^{\text{1PI}}_{k_i\rightarrow 0}
\Bigl\langle \phi_0(k_1+p)\phi_0(-k_1-p)\Bigr\rangle_{k_i\rightarrow 0}\\
&\ \ \times \Bigl\langle \phi_0(k_1+p)\phi_0(k_2)\phi_0(k-p)\Bigr\rangle^{\text{1PI}}_{k_i\rightarrow 0}+(k_1\leftrightarrow k_2)\ . 
\end{aligned}
\end{equation}

In the $C_0^{\text{1PR}}$ expression, the first factor is a three-point functions recognized as the bare Wilson coefficient of the OPE:
\begin{equation}
\phi_0(p)\phi_0(-k_1-p) \sim C_0^1(g_0;p)\phi_0(-k_1)
\end{equation}

The UV divergence is removed by the $Z$-factor:
\begin{equation}
C^1(p) = Z_{\phi}^{\frac{3}{2}} C_0^1(g_0;p) = \text{UV finite}
\end{equation}
The second factor is the exact propagator, which satisfies:
\begin{equation}
Z_{\phi}^{-1}\Bigl\langle \phi_0(k_1+p)\phi_0(-k_1-p)\Bigr\rangle_{k_i\rightarrow 0} = \text{UV finite}
\end{equation}
Since the third factor has the same structure as the first term, we find:
\begin{equation}
Z_{\phi}^{2}C_{0}^{\text{1PR}} = \text{UV finite}
\end{equation}
However, the complete OPE coefficient satisfies:
\begin{equation}
C(p) = \frac{Z_{\phi}}{Z_{\phi^2}}\Bigl[C_0^{\text{1PI}}+C_0^{\text{1PR}}\Bigr] = \text{UV finite}
\end{equation}
Therefore, while $C(p)$ is UV finite, the individual 1PR and 1PI contributions remain divergent.

\begin{figure}[htbp]
	\centering
	\begin{tikzpicture}
		\begin{feynhand}
			\vertex [NEblob] (a0) at (-4.0,0) {};
			\vertex [crossdot] (a1) at (-5.0,-1) {};
			\propag[fer] (a1) to (a0);
			\node at (-4.4,-0.8) {$p$};
			\setlength{\feynhanddotsize}{0.5mm}
			\vertex [dot] (b1) at (-4.2,0.5) {};
			\vertex [dot] (b2) at (-4.0,0.5) {};
			\vertex [dot] (b3) at (-3.8,0.5) {};
			\vertex (a3) at (-5.0,1);
			\propag[chasca] (a3) to (a0);
            \node at (-5.0,0.5) {$k_{1}^{1}$};
            \vertex (c1) at (-3.0,1);
            \propag[chasca] (c1) to (a0);
            \node at (-3.0,0.5) {$k_{1}^{n_{1}}$};
			\vertex [NEblob] (a4) at (-1.0,0) {};
            \vertex [grayblob, style={scale=0.5}] (c2) at (-2.5,0) {};
			\propag[fer] (a0) to (c2);
            \propag[fer] (c2) to (a4);
			\node at (-3.1,-0.35) {$P_{1}$};
            \node at (-1.8,-0.35) {$P_{1}$};
			\vertex (a10) at (0.5,0);
			\propag[fer] (a4) to (a10);
			\node at (0,-0.35) {$P_{2}$};
			\vertex [dot] (b7) at (0.7,0) {};
			\vertex [dot] (b8) at (0.9,0) {};
			\vertex [dot] (b9) at (1.1,0) {};
			\vertex (a11) at (1.3,0);
			\vertex [NEblob] (a5) at (2.8,0) {};
			\propag[fer] (a11) to (a5);
			\node at (1.9,-0.35) {$P_{A-1}$};
            \vertex [grayblob, style={scale=0.5}] (c3) at (4.3,0) {};
			\vertex [NEblob] (a12) at (5.8,0) {};
            \propag[fer] (a5) to (c3);
			\propag[fer] (c3) to (a12);
			\node at (3.6,-0.35) {$P_{A}$};
            \node at (5.0,-0.35) {$P_{A}$};
			\vertex (a6) at (6.8,-1);
			\propag[fer] (a6) to (a12);
			\node at (7.0,-0.55) {$k-p$};
			\vertex [dot] (b4) at (5.6,0.5) {};
			\vertex [dot] (b5) at (5.8,0.5) {};
			\vertex [dot] (b6) at (6.0,0.5) {};
			\vertex (a7) at (4.8,1);
            \vertex (c4) at (6.8,1);
			\propag[chasca] (a7) to (a12);
            \propag[chasca] (c4) to (a12);
            \node at (4.89,0.5) {$k_{A}^{1}$};
            \node at (6.85,0.5) {$k_{A}^{n_{A}}$};
			\vertex (a13) at (-2.0,1);
			\propag[chasca] (a13) to (a4);
            \node at (-2.0,0.5) {$k_{2}^{1}$};
			\vertex (a14) at (0,1);
			\propag[chasca] (a14) to (a4);
            \node at (0,0.5) {$k_{2}^{n_{2}}$};
			\vertex [dot] (b10) at (-1.2,0.5) {};
			\vertex [dot] (b11) at (-1.0,0.5) {};
			\vertex [dot] (b12) at (-0.8,0.5) {};
			\vertex (a15) at (1.8,1);
			\propag[chasca] (a15) to (a5);
            \node at (1.7,0.5) {$k_{A-1}^{1}$};
			\vertex (a16) at (3.8,1);
			\propag[chasca] (a16) to (a5);
            \node at (4.05,0.5) {$k_{A-1}^{n_{A-1}}$};
			\vertex [dot] (b13) at (2.6,0.5) {};
			\vertex [dot] (b14) at (2.8,0.5) {};
			\vertex [dot] (b15) at (3.0,0.5) {};
		\end{feynhand}
	\end{tikzpicture}
	\caption{The chain structure of 1PR diagrams. The crossed dot corresponds to the composite operators in the basis $\Omega_{I}$; the dashed lines denote fields carrying soft momenta, where $K_{i}=\sum_{j}^{n_{i}}k_{i}^{j}$; the gray circle with two external legs denotes exact propagator; every circle overlaid with left slash lines corresponds to a 1PI part. There must be at least one ``soft'' line connected to the end of the chain respectively. \label{fig:chain structure}}
\end{figure}
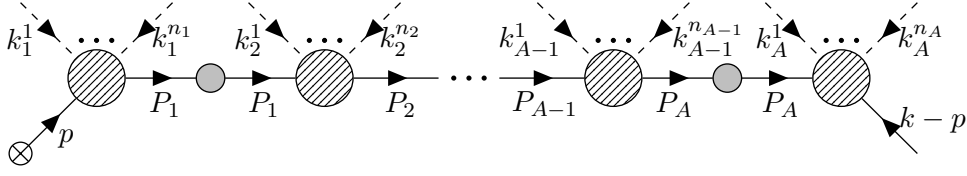 

The hard correlation function can be visualized as a hard particle propagating through a soft background. In a 1PR diagram, its 1PI components must be arranged in a linear chain, as illustrated in Figure~\ref{fig:chain structure}.
The relationship can be expressed as follows:
\begin{equation}
	\begin{aligned}
		&H(p,k-p; k_1,\cdots ,k_n)\\
		=&G^{\text{1PI}}(p,-P_1;K_1)G(P_1)G^{\text{1PI}}(P_1,-P_2;K_2)G(P_2)\cdots G(P_A)G^{\text{1PI}}(P_A,k-p;K_{A+1})\\
		&+\text{all permutations of $k_{i}$} , 
	\end{aligned}
\end{equation}
Each 1PI component possesses two hard external lines and $n_i$ soft external lines, where $n_i\ge1$ and the sum of all $n_i$ equals $n$. These 1PI components are interconnected through exact propagators. Notably, exact propagators do not appear at the chain's boundaries, reflecting the amputation of the hard external legs.

%%%%%%%%%%%%%%%%%%%%%%%
\subsection{Examples of projection operators and OPE coefficients}
%%%%%%%%%%%%%%%%%%%%%%%
We will illustrate how to evaluate projection operators and OPE coefficients through several straightforward examples. The first example considers the dimension-6 operators in $\phi^3$ model, and the second example exhibits the dimension-4 operators in $\phi^{4}$ model. The basis of dimension-6 operators in $\phi^3$ model can be chosen as
\begin{equation}\label{operator basis 1}
    \begin{aligned}
        \Omega_{I}^{0}=\Bigl( \frac{g_{0}^{2}}{3!}\phi^{3}_{0},\ \frac{g_{0}}{2}\partial^{2}\phi^{2}_{0},\ \frac{g_{0}}{2}\phi_{0}\partial^{2}\phi_{0},\ (\partial^{2})^{2}\phi_{0} \Bigr)\ , 
    \end{aligned}
\end{equation}
Here, $g_{0}$ is incorporated into the operator definitions to ensure that the $L$-loop corrections to the hard correlation functions maintain a uniform power counting in $g_{0}$:
\begin{equation}
	\begin{aligned}
		H_{0I}^{(L)}(p;k_{1},\cdots, k_{n})=\mathcal{O}(g_{0}^{n+2L-1})\ . 
	\end{aligned}
\end{equation}
In general, a length-$n$ operator can be expressed as $g_{0}^{n-1}(\partial^{2})^{a}\phi_{0}^{n}$. 

We select the dimension-4 soft operators as 
\begin{equation}
    \begin{aligned}
        O_{\alpha}^{0}=\Bigl( \frac{g_{0}}{2}\phi^{2}_{0},\ \partial^{2}\phi_{0} \Bigr)\ . 
    \end{aligned}
\end{equation}
Their minimal form factors are given by the Feynman rule of vertex $\frac{g_{0}}{2}\phi^{2}_{0}\phi_{0}\phi_{0}$ and $\partial^{2}\phi_{0}\phi_{0}$: 
\begin{equation}
	\begin{aligned}
		F_{1}(k_1,k_2)=g_{0},\ F_{2}(k_1)=-k_{1}^{2}\ . 
	\end{aligned}
\end{equation}
Then by substituting $k_{i}\rightarrow \hat{\partial}_{i}$, 
\begin{equation}
	\begin{aligned}
		F_{1}(\hat{\partial}_1,\hat{\partial}_2)=1,\ F_{2}(\hat{\partial}_1)=-\partial_{1}^{2}\ , 
	\end{aligned}
\end{equation}
and the corresponding inner products $M_{\alpha\beta}$ are 
\begin{equation}
	\begin{aligned}
		(M_{1})_{\alpha\beta}=F_{1}(\hat{\partial}_1,\hat{\partial}_2)F_{1}(k_1,k_2)=g_{0},\ \  (M_{2})_{\alpha\beta}=F_{2}(\hat{\partial}_1)F_{2}(k_1)=-2D\ . 
	\end{aligned}
\end{equation}
The corresponding projection operators are expressed as follows: 
\begin{equation}
	\begin{aligned}
		\mathcal{P}_{02}^{1}=\frac{1}{g_{0}},\ \mathcal{P}_{01}^{2}=-\frac{1}{2D}\hat{\partial}_{1}^{2}\ , 
	\end{aligned}
\end{equation}
where $\hat{\partial}_{1}^{2}=\frac{\partial^{2}}{\partial k_{1}^{2}}$. The OPE coefficients are then determined as: 
\begin{equation}
	\begin{aligned}
		&{C_{0I}}^{1}(p)=\mathcal{P}_{02}^{1}H_{0I}(p;k_{1},k_{2})\Bigr|_{k_{i}=0}=\frac{1}{g_0}H_{0I}(p;k_{1},k_{2})\Bigr|_{k_{i}=0}\ ,  \\
		&{C_{0I}}^{2}(p)=\mathcal{P}_{01}^{2}H_{0I}(p;k_{1})\Bigr|_{k_{i}=0}=-\frac{1}{2D}\hat{\partial}_{1}^{2}H_{0I}(p;k_{1})\Bigr|_{k_{i}=0}\ . 
	\end{aligned}
\end{equation}

Let us analyze the $g_{0}$ dependence of the OPE coefficients. The minimal form factor of a length-$n$ operator $O_{\alpha}^{0}$ scales as
\begin{equation}
	\begin{aligned}
		F_{\alpha}(k_{1},\cdots,k_{n})=\mathcal{O}(g_{0}^{n-1})\ . 
	\end{aligned}
\end{equation}
Consequently, the associated projection operator scales as
\begin{equation}
	\begin{aligned}
		\mathcal{P}_{0n}^{\alpha}=\mathcal{O}(g_{0}^{-n+1})\ . 
	\end{aligned}
\end{equation}
From these relations, we infer that the scaling behavior of OPE coefficients is solely determined by the loop number:
\begin{equation}\label{c-g-power1}
	\begin{aligned}
		{C_{0I}}^{\alpha}(p)=\mathcal{O}(g_{0}^{2L})\ .
	\end{aligned}
\end{equation}

\subsubsection*{The $g_0$ powers in $\phi^4$ model}

The same discussion applies to $\phi^{4}$ theory. Even-length operators can be defined in the form $g_0^{a-1}\partial^{2b}\phi_0^{2a}$, while odd-length operators can be defined in the form $g^{a}\partial^{2b}\phi^{2a+1}$. With this choice of $g_0$ power, $L$-loop hard correlation functions are uniformly 
\begin{equation}
H_{0I}^{(L)}(p;k_1,\cdots, k_n)=\mathcal{O}(g_0^{\lceil\frac{n+1}{2}\rceil+L-1})\ ,
\end{equation} 
where $\lceil\rceil$ is the ceiling function.

For example, the basis of dimension-4 operators in $\phi^{4}$ model (even-length) can be chosen as
\begin{equation}
\Omega_I^0=\Bigl(\frac{1}{4!}g_0\phi_0^4,\ \frac{1}{2}\partial^2\phi_0^2,\  \phi_0\partial^2\phi_0\Bigr)\ .
\end{equation}
And the basis of dimension-3 scalar operators (odd-length) can be chosen as
\begin{equation}
O_{\alpha}^0=\Bigl(\frac{1}{3!}g_0\phi_0^3,\ \partial^2\phi_0\Bigr)\ .
\end{equation}

The minimal form factor of a length-$n$ operator $O_{\alpha}^0$ is of the order
\begin{equation}
F_{\alpha}(k_1,\cdots, k_n)=\mathcal{O}(g_0^{\lceil\frac{n}{2}\rceil-1})\ .
\end{equation}
Then the corresponding projector is of the order
\begin{equation}
\mathcal{P}_{0n}^{\alpha}=\mathcal{O}(g_0^{-\lceil\frac{n}{2}\rceil+1})\ .
\end{equation}
The OPE coefficient is of the order
\begin{equation}\label{c-g-power}
C_{0I}^{\ \ \alpha}(p)=\mathcal{O}(g_0^{\lceil\frac{n+1}{2}\rceil-\lceil\frac{n}{2}\rceil+L})\ .
\end{equation}

Thus, when $\Omega_I^0$ are even-length operators, the OPE coefficients acquire an extra factor of $g_0$. This additional factor arises from the inherent difference in operator definitions between the even-length basis $\Omega_I^0$ and the odd-length basis $O_{\alpha}^0$.

%%%%%%%%%%%%%%%%%%%%%%%%%%%%%%%%%%%%%%%%%%%
\section{The basis of soft operators}\label{selecting O_alpha}
%%%%%%%%%%%%%%%%%%%%%%%%%%%%%%%%%%%%%%%%%%%

As discussed in Section \ref{ope-o-phi}, the UV finiteness of ${C_{I}}^{\alpha}$ imposes constraints on ${Z_{I}}^{J}$ via the relation \eqref{ope-coe-relation}. We must select a set of soft operator $O_{\alpha}$ whose $Z$-factors are more tractable, while still sufficient to fully determine ${Z_{I}}^{J}$. In this section, we establish the rule for selecting a sufficient set of soft operator. The critical requirement is that the matrix formed by their tree-level OPE coefficients possesses full row rank, ensuring the system of equations derived from UV finiteness is non-degenerate and solvable. We demonstrate that a set of lower-dimensional symmetric traceless tensors satisfies this criterion. These tensor operators are either descendants (sharing $Z$-factors with primaries), conserved currents, or can be combined with total derivatives to form scalar operators which are simpler than $\Omega_I$. By utilizing this framework, we establish a recursive procedure. It allows us to determine the anomalous dimensions for all scalar operators starting from lower scaling dimensions and proceeding to higher ones.

%%%%%%%%%%%%%%%%%%%%%%%%%%%%%%%%%%%%%%%%%%%
\subsection{The criteria to determine the soft operator basis}
%%%%%%%%%%%%%%%%%%%%%%%%%%%%%%%%%%%%%%%%%%%

To establish the criteria for a sufficient soft operator basis, we analyze the structure of the OPE coefficients in $\phi^4$ model at different loop orders.
Let us expand the quantities in \eqref{ope-coe-relation} to 1-loop order: 
\begin{equation}\label{ope-expand-loop}
    \begin{aligned}
        {C_{I}}^{\alpha}=&g^{\delta}{\Bigl(C^{(0)}\Bigr)_{I}}^{\alpha}+g^{\delta+1}{\Bigl(C^{(1)}\Bigr)_{I}}^{\alpha}+\mathcal{O}(g^{\delta+2})\ ,  \\
        {Z_{I}}^{J}=&\delta_{I}^{J}+g{\Bigl(Z^{(1)}\Bigr)_{I}}^{J}+\mathcal{O}(g^{2})\ ,\\
        {(C_{0})_{J}}^{\beta}=&g^{\delta}{\Bigl(C_{0}^{(0)}\Bigr)_{J}}^{\beta}+g^{\delta+1}{\Bigl(C_{0}^{(1)}\Bigr)_{J}}^{\beta}+\mathcal{O}(g^{\delta+2})\ ,  \\
        {(Z^{-1})_{\beta}}^{\alpha}=&\delta_{\beta}^{\alpha}-g{\Bigl(Z^{(1)}\Bigr)_{\beta}}^{\alpha}+\mathcal{O}(g^{2})\ . 
    \end{aligned}
\end{equation}
The extra $g^{\delta}$ factor in the OPE coefficients was discussed in \eqref{c-g-power}. At $\mathcal{O}(g^{\delta})$ (tree level), the matching condition is 
\begin{equation}
    \begin{aligned}
        {\Bigl(C^{(0)}\Bigr)_{I}}^{\alpha}={\Bigl(C_{0}^{(0)}\Bigr)_{I}}^{\alpha}\ . 
    \end{aligned}
\end{equation}
At $\mathcal{O}(g^{\delta+1})$ (one-loop order), the matching condition yields a linear system for the matrix ${\Bigl(Z^{(1)}\Bigr)_{I}}^{J}$: 
\begin{equation}\label{Z-determine-1-loop}
    \begin{aligned}
        &-{\Bigl(C^{(0)}\Bigr)_{I}}^{\beta}{\Bigl(Z^{(1)}\Bigr)_{\beta}}^{\alpha}+{\Bigl(C_{0}^{(1)}\Bigr)_{I}}^{\alpha}+{\Bigl(Z^{(1)}\Bigr)_{I}}^{J}{\Bigl(C^{(0)}\Bigr)_{J}}^{\alpha}=\text{UV finite}\ . 
    \end{aligned}
\end{equation}
For this linear system to have a unique solution for ${\Bigl(Z^{(1)}\Bigr)_{I}}^{J}$, the matrix formed by the tree-level OPE coefficients must possess full row rank. This ensures that the system is non-degenerate and invertible. Let $N_{\Omega}$ and $N_{O}$ denote the number of operators in $\Omega_{I}$ and $O_{\alpha}$, respectively. The critical criterion is:
\begin{equation}\label{choose-operator-1}
    \begin{aligned}
        \text{rank}{\Bigl(C^{(0)}\Bigr)_{J}}^{\alpha}=N_{\Omega}\ . 
    \end{aligned}
\end{equation}
The same condition ensures that the $L$-loop correction of ${Z_{I}}^{J}$ can be determined from the $L$-loop OPE coefficients, as can be proved by expanding \eqref{ope-expand-loop} to $\mathcal{O}(g^{\delta+L})$. 

As an example, let us continue with our discussion about the dimension-4 scalar operators in $\phi^4$ model and the soft basis $O_{\alpha}^{0}$ has been chosen as following: 
\begin{equation}\label{choose-op-example}
    \begin{aligned}
        {\Omega_{I}}^{0}=&\Bigl(\frac{g_{0}}{4!}\phi_{0}^{4},\ \frac{1}{2}\partial^{2}\phi_{0}^{2},\ \phi_{0}\partial^{2}\phi_{0}\Bigr)\ ,  \\
        O_{\alpha}^{0}=&\Bigl(\phi_{0},\ \partial_{\mu}\phi_{0},\ \frac{g_{0}}{3!}\phi_{0}^{3},\ \partial^{2}\phi_{0}\Bigr)\ . 
    \end{aligned}
\end{equation}

The minimal form factors are 
\begin{equation}
    \begin{aligned}
        F_{01}(k_{1})=1\ ,\ F_{02}(k_{1})=ik_{1\mu}\ ,\ F_{03}(k_{1},k_{2},k_{3})=g_{0}\ ,\ F_{04}(k_{1})=-k_{1}^{2}\ . 
    \end{aligned}
\end{equation}

The projection operators are 
\begin{equation}\label{projection operators1}
    \begin{aligned}
        \mathcal{P}_{01}^{1}=1\ ,\ 
        \mathcal{P}_{01}^{2}=-i\hat{\partial}_{1}^{\mu}\ ,\ 
        \mathcal{P}_{03}^{3}=\frac{1}{g_{0}}\ ,\ 
        \mathcal{P}_{01}^{4}=-\frac{1}{2D}\hat{\partial}_{1}^{2}\ . 
    \end{aligned}
\end{equation}

The tree-level hard correlation functions are 
\begin{equation}
    \begin{aligned}
        H_{0I}^{(0)}(p;k_{1},k_{2},k_{3})=&\Bigl(g_{0},\ \sum_{i=1}^3\frac{-g_{0}p^{2}}{(p+k_{i})^{2}},\ \sum_{i=1}^{3}\frac{-g_{0}[k_{i}^{2}+(p+k_{i})^{2}]}{(p+k_{i})^{2}}\Bigr)\ ,\\
        H_{0I}^{(0)}(p;k_{1})=&\Bigl(0,-p^{2},-k_{1}^{2}-(p+k_{1})^{2}\Bigr)\ . 
    \end{aligned}
\end{equation}

The tree-level bare OPE coefficients are 
\begin{equation}
    \begin{aligned}
        {\Bigl(C_0^{(0)}\Bigr)_{I}}^{1}=&\mathcal{P}_{01}^{1}H_{0I}^{(0)}(p;k_{1})\Bigr|_{k_{1}=0}=\Bigl(0,-p^{2},-p^{2}\Bigr)\ ,\\
        {\Bigl(C_0^{(0)}\Bigr)_{I}}^{2}=&\mathcal{P}_{01}^{2}H_{0I}^{(0)}(p;k_{1})\Bigr|_{k_{1}=0}=\Bigl(0,0,2ip^{\mu}\Bigr)\ ,\\
        {\Bigl(C_0^{(0)}\Bigr)_{I}}^{3}=&\mathcal{P}_{03}^{3}H_{0I}^{(0)}(p;k_{1},k_{2},k_{3})\Bigr|_{k_{i}=0}=\Bigl(1,-3,-3\Bigr)\ ,\\
        {\Bigl(C_0^{(0)}\Bigr)_{I}}^{4}=&\mathcal{P}_{01}^{4}H_{0I}^{(0)}(p;k_{1})\Bigr|_{k_{1}=0}=\Bigl(0,0,2\Bigr)\ . 
    \end{aligned}
\end{equation}
Since the rank of the matrix ${\Bigl(C_{0}^{(0)}\Bigr)_{I}}^{\alpha}$ equals to $N_{\Omega}$, the OPE coefficients corresponding to \eqref{choose-op-example} are sufficient to determine the $Z$ factors of $\Omega_{I}^{0}$ in \eqref{choose-op-example}. In fact, ${\Bigl(C_{0}^{(0)}\Bigr)_{I}}^{\alpha}$ is redundant, and $(O^{0}_{1},\ O^{0}_{2},\ O^{0}_{3})$ or $(O^{0}_{1},\ O^{0}_{3},\ O^{0}_{4})$ are both sufficient to determine ${Z_{I}}^{J}$. But since $O^{0}_{3}$ and $O^{0}_{4}$ are mixed, which means the renormalization of $O^{0}_{3}$ and $O^{0}_{4}$ shall be considered together, it is more convenient to choose $(O^{0}_{1},\ O^{0}_{3},\ O^{0}_{4})$. 

%%%%%%%%%%%%%%%%%%%%%%%%%%%%%%%%%%%%%%%%%%%
\subsection{The soft operator basis for generic scalar operators}
%%%%%%%%%%%%%%%%%%%%%%%%%%%%%%%%%%%%%%%%%%%

We now consider $\Omega_I$ for generic $\Delta$. We begin by focusing on $\Omega_{I}$ with fixed lengths, which takes the form $(\partial^{2})^{s}\phi^{n+1}$.
We demonstrate that the soft operators $O_{\alpha}$ of the structure $(\partial^{2})^{r}\partial_{(\mu_{1}\cdots\mu_{J})}\phi^{n}$ are sufficient to determine the $Z$-factors of $\Omega_{I}$, where $\partial_{(\mu_{1}\cdots\mu_{J})}$ denotes the symmetric traceless component of $\partial_{\mu_{1}}\cdots\partial_{\mu_{J}}$, and $r$ and $J$ run through non-negative integers satisfying
\begin{equation}\label{spin-constraint}
    \begin{aligned}
        r+J\le s\ . 
    \end{aligned}
\end{equation}

To illustrate the mechanism, we first consider the specific case where $n=1$ and $s=2$. The hard operators $\Omega_{I}$ take the form:
\begin{equation}\label{op-n1s2}
    \begin{aligned}
        \Omega_{I}=\Bigl(\phi(\partial^{2})^{2}\phi,\ \frac{1}{2}(\partial^{2}\phi)^{2},\ \partial^{2}(\phi\partial^{2}\phi),\ \frac{1}{2}(\partial^{2})^{2}\phi^{2}\Bigr)\ . 
    \end{aligned}
\end{equation}
The corresponding soft operator basis $O_{rJ}$, constructed from  symmetric traceless tensors with $r+J\le s=2$, is given by: 
\begin{equation}\label{all-o}
    \begin{aligned}
        O_{\alpha}=&O_{00}\cup O_{01}\cup O_{10}\cup O_{02}\cup O_{11}\cup O_{20}\\
        =&\Bigl(\phi,\ \partial^{\mu}\phi,\ \partial^{2}\phi,\ \partial^{(\mu_{1}\mu_{2})}\phi,\ \partial^{\mu}\partial^{2}\phi,\ (\partial^{2})^{2}\phi\Bigr)\ .
    \end{aligned}
\end{equation}

The tree-level hard correlation functions associated with \eqref{op-n1s2} are 
\begin{equation}
    \begin{aligned}
        H_{I}^{(0)}=\Bigl((k_{1}^{2})^{2}+[(p+k_{1})^{2}]^{2},\ k_{1}^{2}(p+k_{1})^{2},\ p^{2}[k_{1}^{2}+(p+k_{1})^{2}],\ (p^{2})^{2}\Bigr)\ , 
    \end{aligned}
\end{equation}
and the minimal form factors corresponding to \eqref{all-o} are 
\begin{equation}
    \begin{aligned}
    F_{\alpha}=\Bigl(1,\ ik_{1}^{\mu},\ -k_{1}^{2},\ -k_{1}^{(\mu_{1}}k_{1}^{\mu_{2})},\ -ik_{1}^{\mu}k_{1}^{2},\ (k_{1}^{2})^{2}\Bigr)\ , 
    \end{aligned}
\end{equation}
By matching these polynomials, we construct the tree-level OPE coefficients matrix:
\begin{equation}\label{ope-coe-example}
    \begin{aligned}
        {\Bigl(C^{(0)}\Bigr)_{I}}^{\alpha}=
        \begin{pNiceMatrix}[first-row, first-col]
                & \textcolor{gray!70}{\phi}\ & \textcolor{gray!70}{\partial^{\mu}\phi}\ & \textcolor{gray!70}{\partial^{2}\phi}\ & \textcolor{gray!70}{\partial^{(\mu_{1}\mu_{2})}\phi}\ & \textcolor{gray!70}{\partial^{\mu}\partial^{2}\phi}\ & \textcolor{gray!70}{(\partial^{2})^{2}\phi}  \\
                \textcolor{gray!70}{\phi(\partial^{2})^{2}\phi}\ & (p^{2})^{2}\ & -4ip^{2}p_{\mu}\ & \frac{2(2-D)}{D}p^{2}\ & -4p_{\mu_{1}}p_{\mu_{2}}\ & 4ip_{\mu}\ & 2  \\
                \textcolor{gray!70}{\frac{1}{2}(\partial^{2}\phi)^{2}}\ & 0\ & 0\ & -p^{2}\ & 0\ & 2ip_{\mu},\ &1  \\
                \textcolor{gray!70}{\partial^{2}(\phi\partial^{2}\phi)}\ & (p^{2})^{2}\ & -2ip^{2}p_{\mu}\ & -2p^{2}\ & 0\ & 0\ &0  \\
                \textcolor{gray!70}{\frac{1}{2}(\partial^{2})^{2}\phi^{2}}\ & (p^{2})^{2}\ & 0\ &0\ & 0\ & 0\ &0  \\
        \end{pNiceMatrix}\ . 
    \end{aligned}
\end{equation}
The matrix \eqref{ope-coe-example} has full row rank, confirming that the chosen soft basis is sufficient for this specific case. 

For general $s$ and $n$, the tree-level hard correlation functions are degree-$2s$ polynomials of Mandelstam variables constructed from $p$ and $k_{i}$: 
\begin{equation}
    \begin{aligned}
        H_{0I}^{(0)}\in \mathbb{R}\Bigl[p^{2}, p\cdot k_{i}, k_{i}^{2}, k_{i}\cdot k_{j}\Bigr],\ \text{deg}H_{0I}^{(0)}=2s\ . 
    \end{aligned}
\end{equation}
In fact, $H_{0I}^{(0)}$ form a basis of degree-$2s$ polynomials with $S_{n+1}$ symmetry. This $S_{n+1}$ symmetry is generated by the permutations of $(k-p,\ k_{1},\cdots, k_{n})$. 

The minimal form factors, on the other hand, are polynomials constructed from $k_i$ alone:
\begin{equation}
    \begin{aligned}
        F_{\alpha}(k_{1},\cdots, k_{n})\in \mathbb{R}\Bigl[k_{i}^{2}, k_{i}\cdot k_{j}, k_{i}^{\mu_{a}}\Bigr]\ . 
    \end{aligned}
\end{equation}
The tree-level OPE coefficients can be derived by matching these polynomials: 
\begin{equation}\label{OPE-match}
    \begin{aligned}
        H_{0I}^{(0)}(p;k_{1},\cdots,k_{n})={\Bigl(C_{0}^{(0)}\Bigr)_{I}}^{\alpha}(p)F_{\alpha}(k_{1},\cdots,k_{n})\ , 
    \end{aligned}
\end{equation}
where the Lorentz indices in each minimal form factor are contracted with those in the corresponding OPE coefficient. We observe that the OPE coefficients can be factored as: 
\begin{equation}\label{OPE-coe-factor}
    \begin{aligned}
        {\Bigl(C_{0}^{(0)}(p)\Bigr)_{I}}^{\alpha}={\Bigl(c_{0}^{(0)}\Bigr)_{I}}^{\alpha}(p^{2})^{s-r_{\alpha}-J_{\alpha}}p^{\mu_{1}}\cdots p^{\mu_{J_{\alpha}}}\ . 
    \end{aligned}
\end{equation}

We now combine these $p$-dependent terms with the corresponding minimal form factors into a single polynomial, denoted $\hat{F}_{\alpha}$. The remaining terms in the OPE coefficients, denoted $c_{0}^{(0)}$, are constants. Substituting \eqref{OPE-coe-factor} into \eqref{OPE-match} gives: 
\begin{equation}\label{OPE-match-1}
    \begin{aligned}
        H_{0I}^{(0)}(p;k_{1},\cdots,k_{n})={\Bigl(c_{0}^{(0)}\Bigr)_{I}}^{\alpha}\hat{F}_{\alpha}(p^{2},p\cdot k_{i},k_{i}^{2},k_{i}\cdot k_{j})\ . 
    \end{aligned}
\end{equation}
Notably, $\hat{F}_{\alpha}$ also form a basis of degree-$2s$ polynomials constructed from $p$ and $k_{i}$, but they possess $S_{n}$ symmetry (generated by the permutations of $(k_{1},\cdots,k_{n})$). The linear space spanned by $H_{0I}^{(0)}$ is therefore a subset of that spanned by $\hat{F}_{\alpha}$, since the former preserves a larger symmetry group than the latter. As a result, the matrix ${\Bigl(c_{0}^{(0)}\Bigr)_{I}}^{\alpha}$, and consequently ${\Bigl(C_{0}^{(0)}\Bigr)_{I}}^{\alpha}$, has full row rank. 

In general case, operators with different lengths mix with one another, as can be seen from the example in \eqref{choose-op-example}. Let us first classify the dimension-$\Delta$ operators $\Omega_{I}$ by their lengths, which denoted by $\mathbb{L}$, 
\begin{equation}\label{hard-choice}
    \begin{aligned}
        \Omega_{I}=\Omega_{I}^{\mathbb{L}}\cup \Omega_{I}^{\mathbb{L}-2}\cup \Omega_{I}^{\mathbb{L}-4}\cup\cdots\ , 
    \end{aligned}
\end{equation}
in which operators in $\Omega_{I}^{\mathbb{L}}$ are of the form $(\partial^{2})^{\frac{\Delta-\mathbb{L}}{2}}\phi^{\mathbb{L}}$. 

The soft operators can still be chosen in the same manner as the fixed length case: 
\begin{equation}\label{soft-choice}
    \begin{aligned}
        O_{\alpha}=O_{\alpha}^{\mathbb{L}-1}\cup O_{\alpha}^{\mathbb{L}-3}\cup O_{\alpha}^{\mathbb{L}-5}\cup \cdots\ , 
    \end{aligned}
\end{equation}
in which $O_{\alpha}^{\mathbb{L}-1}$ are of the structure $(\partial^{2})^{r}\partial_{(\mu_{1}\cdots\mu_{J})}\phi^{\mathbb{L}-1}$ satisfying 
\begin{equation}
    \begin{aligned}
        r+J\le \frac{\Delta-\mathbb{L}}{2}\ , 
    \end{aligned}
\end{equation}
which correspond to the hard operators $(\partial^{2})^{\frac{\Delta-\mathbb{L}}{2}}\phi^{\mathbb{L}}$ in the fixed length case. 

\begin{table}[H]
\centering
\begin{tabular}{|c |c|c|c|c|c|}
\hline
\diagbox{$\mathbb{L}(\Omega_I)$}{$\mathbb{L}(O_{\alpha})$} & $\mathbb{L}-1$ & $\mathbb{L}-3$  &  $\mathbb{L}-5$ & $\cdots$   \\
\hline
$\mathbb{L}$ & $\ne 0$ & $0$  &  $0$ & $0$   \\
\hline
$\mathbb{L}-2$ & $\ne 0$ & $\ne0$  &  $0$ & $0$   \\
\hline
$\mathbb{L}-4$ & $\ne 0$ & $\ne0$  &  $\ne0$ & $0$   \\
\hline
$\cdots$ & $\ne 0$ & $\ne0$  &  $\ne0$ & $\ne0$   \\
\hline
\end{tabular}
\caption{The matrix ${\Bigl(C_{0}^{(0)}\Bigr)_{I}}^{\alpha}$ is block-triangular. }
\label{block-triangular}
\end{table}

In order to prove that $O_{\alpha}$ in \eqref{soft-choice} are sufficient to determine the $Z$-factors of $\Omega_{I}$ in \eqref{hard-choice}, we observe that the matrix ${\Bigl(C_{0}^{(0)}\Bigr)_{I}}^{\alpha}$, as demonstrated in Table \ref{block-triangular}, is block-triangular. The tree-level hard correlation function $\mathcal{H}_{I}(p;k_{1},\cdots,k_{n})$ vanishes if 
\begin{equation}
    \begin{aligned}
        \mathbb{L}(\Omega_{I})> \mathbb{L}(O_{\alpha})+1\ . 
    \end{aligned}
\end{equation}
Consequently, the corresponding ${\Bigl(C_{0}^{(0)}\Bigr)_{I}}^{\alpha}$ also vanishes because it is obtained by acting the projection operator on the hard correlation function. For example, in the first row of Table \ref{block-triangular}, only the first entry is non-vanishing, where 
\begin{equation}
    \begin{aligned}
        \mathbb{L}(\Omega_{I})=\mathbb{L}(O_{\alpha})+1\ . 
    \end{aligned}
\end{equation}
Since each of its diagonal block has full row rank, ${\Bigl(C_{0}^{(0)}\Bigr)_{I}}^{\alpha}$ also has full row rank. 

%%%%%%%%%%%%%%%%%%%%%%%%%%%%%%%%%%%%%%%%%%%
\subsection{The renormalization of tensor operators}
%%%%%%%%%%%%%%%%%%%%%%%%%%%%%%%%%%%%%%%%%%%
To date, our discussion has been restricted to the case where $\Omega_{I}$ are scalar operators. While this analysis can be extended to the scenarios where both $\Omega_{I}$ and $O_{\alpha}$ are tensor operator belonging to more complicated representations of $SO(1,D-1)$ (corresponding to Young tableaux with more than one row), this is not the focus of the current paper. We must, however, investigate the renormalization of symmetric traceless tensors to ensure the algorithm closes. Such tensors emerge as soft operators on the right-hand side of OPE, even when $\Omega_{I}$ are scalar operators. 

There are several ways to evaluate the $Z$-factors of symmetric traceless tensor operators. Some tensor operators are descendants (total derivatives of other operators), and their $Z$-factors are the same as the corresponding primaries. We may therefore focus on the primary tensor operators. 

Tensor operators can be combined with total derivatives to form scalar operators. The $Z$-factors of the tensor operators are the same as the resulting scalar operators, unless the latter vanishes (i.e., when the tensor operators are conserved currents). The soft operators take the form $(\partial^{2})^{r}\partial_{(\mu_{1}\cdots\mu_{J})}\phi^{n}$, so the resulting scalar operators are of the form $(\partial^{2})^{r+J}\phi^{n}$, with dimension $\Delta=2r+2J+n$. Using \eqref{spin-constraint}, this dimension remains less than that of $\Omega_{I}$: 
\begin{equation}
    \begin{aligned}
        2r+2J+n< 2s+n+1\ , 
    \end{aligned}
\end{equation}
meaning the renormalization of these resulting operators is still simpler than that of $\Omega_{I}$. 

The only tensor operators which cannot be computed via these two methods are conserved currents, which are not renormalized. In the $\phi^{4}$ model, the only conserved currents are the energy stress tensor $T_{\mu\nu}$ and its descendants~\cite{Callan:1970ze,Collins:1976vm,Peskin:1995ev}. 
%In the free field limit, higher spin currents are also conserved, and $T_{\mu\nu}$ can be treated as a special case of higher spin currents. Since $T_{\mu\nu}$ is a conserved current, it is not renormalized. Higher spin currents, meanwhile, are conserved in the free field limit, and their anomalous dimensions are known up to 4-loops.

Combining these techniques allows us to compute the anomalous dimensions of all scalar operators in the $\phi^{4}$ model. Let us use the $\partial_{(\mu_{1}\mu_{2})}\phi^{2}$ type operators as an example. The operator basis can be chosen as 
\begin{equation}
    \begin{aligned}
        \Omega_{I}=\Bigl(\frac{1}{2}\partial_{(\mu_{1}\mu_{2})}\phi^{2},\ \partial_{(\mu_{1}}\phi\partial_{\mu_{2})}\phi\Bigr)\ , 
    \end{aligned}
\end{equation}
in which, $\Omega_{1}$ is a descendant operator, while $\Omega_{2}$ can be contracted with total derivatives: 
\begin{equation}
    \begin{aligned}
        \partial^{\mu_{1}}\partial^{\mu_{2}}[\partial_{(\mu_{1}}\phi\partial_{\mu_{2})}\phi]=-\frac{1}{4}\phi(\partial^{2})^{2}\phi+\frac{1}{2}(\partial^{2}\phi)^{2}+\frac{1}{2D}\partial^{2}(\phi\partial^{2}\phi)+\frac{D-2}{4D}(\partial^{2})^{2}\phi^{2}\ , 
    \end{aligned}
\end{equation}
which is a linear combination of $(\partial^{2})^{2}\phi^{2}$ type operators in \eqref{op-n1s2}. 

%\textcolor{red}{new subsection on recursion}

%%%%%%%%%%%%%%%%%%%%%%%%%%%%%%%%%%%%%%%%%%%
\section{Renormalization of operators in $\phi^3$ model}\label{demonstration}
%%%%%%%%%%%%%%%%%%%%%%%%%%%%%%%%%%%%%%%%%%%
In this section, we present the computational details of renormalizing mixed composite operators within the $\phi^{3}$ model. The analysis begins with operators of the lowest dimension, and we determine anomalous dimensions recursively, progressing from lower to higher dimensions. Our calculations encompass the anomalous dimensions of operators ranging from dimension-2 to dimension-10, extending up to two-loop order. Additionally, we discuss descendant operators, which facilitate algorithmic optimization. 

%For a set of operators $\Omega_{I}^{\Delta}$ with dimension $\Delta$, the maximal length of operators in $\Omega_{I}^{\Delta}$ is $Q_{\text{max}} = \frac{\Delta}{2}$. When computing hard correlation functions, operators of length $Q$ always mix with operators of length $Q^{\prime} > Q$. However,  operators of length $Q^{\prime}$ mix with operators of length $Q$, subject to the condition $(Q^{\prime}- L) \leq Q \leq Q^{\prime}$, where $L$ denotes the number of loops in the Feynman diagram (exhibited in Table \ref{ext-length}). Consequently, it suffices to consider hard correlation functions of length-$Q^{\prime}$ operators whose number of external legs ranges from $Q^{\prime}- L$ to $Q_{\text{max}}$. 

%%%%%%%%%%%%%%%%%%%%%%%%%%%%%%%%%%%%%%%%%%%
\subsection{Field anomalous dimension and the beta function}
%%%%%%%%%%%%%%%%%%%%%%%%%%%%%%%%%%%%%%%%%%%
The fundamental field $\phi$ is the sole dimension-2 operator in the theory. Its anomalous dimension is determined from the two-point correlation function, which is 
\begin{equation}
    \gamma_{\phi}=\frac{g^{2}}{12}+\frac{13 g^{4}}{432}\ .
\end{equation}

The beta function is derived by employing the OPE of two fundamental fields:
\begin{equation}
    \phi(k-p)\phi(p)\sim C(p)\phi(k)\ .
\end{equation}
The bare OPE coefficient can be extracted from the 3-point correlation function by setting one external momentum to zero:
\begin{equation}
    C_0(g_0;p)=\Bigl\langle  \phi_0(k-p)\phi_0(p)\phi_0(-k)\Bigr\rangle^{\text{amp}}_{k\rightarrow0}\ .
\end{equation}
The relationship between the renormalized and bare OPE coefficients is given by
\begin{equation}
    C(p)=Z_{\phi}^{\frac{3}{2}}C_0(g_0;p)\ .
\end{equation}
Requiring the OPE coefficient $C(p)$ to be UV finite determines the pole part of the coupling renormalization constant $Z_{g}$ order by order. The beta function $\beta(g)$ then follows from the relation between the bare and renormalized couplings. 
%The UV-finiteness of $C(p)$ uniquely fixes $g_0$, and from which the beta function $\beta(g)$ can be determined: 
\begin{equation}
    \frac{\partial g}{\partial \ln \mu}=-\epsilon g-\frac{3g^{3}}{4}-\frac{125 g^{5}}{144}\ .
\end{equation}

%%%%%%%%%%%%%%%%%%%%%%%%%%%%%%%%%%%%%%%%%%%
\subsection{Dimension-4 operators}
%%%%%%%%%%%%%%%%%%%%%%%%%%%%%%%%%%%%%%%%%%%
The emergence of composite operators begins at the dimension of four. The corresponding operator basis is defined as follows:
\begin{equation}\label{operator basis 2}
    \begin{aligned}
        \Omega_{I}^{0}=\Bigl( \frac{g_{0}}{2}\phi^{2}_{0},\ \partial^{2}\phi_{0} \Bigr)\ , 
    \end{aligned}
\end{equation}
and the soft basis is chosen as 
\begin{equation}\label{phi3-soft-basis-1}
    \begin{aligned}
        O_{\alpha}^{0}=\Bigl( I,\ \phi_{0} \Bigr)\ . 
    \end{aligned}
\end{equation}

The following hard correlation functions are required:
\begin{equation}
\begin{aligned}
H_{I0}(p)=&\Bigl\langle \Omega_I(p)\phi(-p)\Bigr\rangle_h\ ,\\
H_{I0}(p;k_1)=&\Bigl\langle \Omega_I(p)\phi(-k_1-p)\phi(k_1)\Bigr\rangle_h\ .\\
\end{aligned}
\end{equation}
The projection operators corresponding to \eqref{phi3-soft-basis-1} are simply
\begin{equation}
    P_{00}^{1}=P_{01}^{2}=1\ .
\end{equation}
Applying the projection operators to the hard correlation functions, the bare OPE coefficients are obtained:
\begin{equation}
\begin{aligned}
C_{0I}^1(p)=&P_{00}^{1}H_{I0}(p)=H_{I0}(p)\ ,\\
C_{0I}^2(p)=&P_{01}^{2}H_{I0}(p;k_1)=H_{I0}(p;0)\ .\\
\end{aligned}
\end{equation}

Two soft operators in \eqref{phi3-soft-basis-1} are not mixed, so their OPE coefficients can be evaluated independently. For the first soft operator $I$, the inverse Z-matrix $(Z^{-1})_{\alpha}^{\ \beta}=1$, and we obtain:
\begin{equation}
    \begin{aligned}
        &Z_{\phi}^{\frac{1}{2}}Z_I^{\ J}C_{0J}^1(p)
        =\text{UV-finite}\ . \\
    \end{aligned}\label{uv-finite-41}
\end{equation}
For the second soft operator $\phi_0$, the inverse Z-matrix 
\begin{equation}
    (Z^{-1})_{\alpha}^{\ \beta}=Z_{\phi}^{\frac{1}{2}}\ ,
\end{equation}
and we obtain:
\begin{equation}
    \begin{aligned}
        &Z_{\phi}^{\frac{1}{2}}Z_I^{\ J}C_{0J}^2(p)Z_{\phi}^{\frac{1}{2}}
        =\text{UV-finite}\ .
    \end{aligned}\label{uv-finite-42}
\end{equation}

These two conditions, \eqref{uv-finite-41} and\eqref{uv-finite-42}, uniquely fix the Z-matrix. The expression to two-loop is given by: 
\begin{equation}\label{renormalization-1}
    \begin{aligned}
        {Z_{I}}^{J}=
        \begin{pNiceMatrix}[first-row]
            \textcolor{gray!70}{\frac{g_{0}}{2}\phi^{2}_{0}} & \textcolor{gray!70}{\partial^{2}\phi_{0}} \\
            1-\frac{g^{2}}{24\epsilon}-\frac{13g^{4}}{1728\epsilon}+\frac{19g^{4}}{1152\epsilon^{2}} & \frac{g^{2}}{12\epsilon}+\frac{13g^{4}}{864\epsilon}-\frac{g^{4}}{32\epsilon^{2}} \\
            0 & 1+\frac{g^{2}}{24\epsilon}+\frac{13g^{4}}{1728\epsilon}-\frac{17g^{4}}{1152\epsilon^{2}}
        \end{pNiceMatrix}\ . 
    \end{aligned}
\end{equation}
The anomalous dimension ${\gamma_{I}}^{J}$ is determined by 
\begin{equation}
    \begin{aligned}
        {\gamma_{I}}^{J}=-\frac{\partial {Z_{I}}^{K}}{\partial\ln\mu}{(Z^{-1})_{K}}^{J}=
        \begin{pNiceMatrix}
            -\frac{g^{2}}{12}-\frac{13 g^{4}}{432} \ \ \ & -\frac{g^{2}}{6}-\frac{13 g^{4}}{216}\\
            0 & \frac{g^{2}}{12}+\frac{13 g^{4}}{432}
        \end{pNiceMatrix}\ . 
    \end{aligned}\label{gamma-dim-4}
\end{equation}

The eigenvalues of (\ref{gamma-dim-4}) are both proportional to the anomalous dimension of $\phi$. This is not surprising, because these dimension-4 operators are descendant operators. In the next subsection we review the renormalization of descendant operators. The knowledge of descendant operators helps us to optimize the method, and at the same time provides a consistent check to our computations.

%%%%%%%%%%%%%%%%%%%%%%%%%%%%%%%%%%%%%%%%%%%
\subsection{Renormalization of descendant operators}\label{re-descedant}
%%%%%%%%%%%%%%%%%%%%%%%%%%%%%%%%%%%%%%%%%%%
The OPE method is applicable to both primary and descendant operators, where descendant operators can be divided into two types: the total derivative operators and the equation of motion (EOM) operators (i.e. operators proportional to the EOM). The renormalization properties of descendant operators are non-independent, and can be determined from the renormalization of lower-dimensional operators. The OPE method can be optimized by incorporating as many descendant operators as possible in the operator basis. 

The renormalization of total derivative operators is relatively straightforward, as their renormalization behavior is identical to that of their corresponding primary operators. Here, we provide some details on EOM operators. 

We denote the equation of motion as $E$, and analyze the operators proportional to $E$, which we hereafter refer to as EOM operators for brevity. In the $\phi^3$ model, $E$ is expressed as: 
\begin{equation}
    \begin{aligned}
        E=Z_{\phi}\partial^{2}\phi+\frac{Z_gg\tilde{\mu}^{\epsilon}}{2}\phi^2=Z_{\phi}^{\frac{1}{2}}\Bigl(\partial^2\phi_0+\frac{g_0}{2}\phi_0^{2}\Bigr)=Z_{\phi}^{\frac{1}{2}}E_0\ .
    \end{aligned}
\end{equation}

For a renormalized operator $\Omega_I^R$, the correlation functions of $\Omega_I^R E$ can be reduced using the Schwinger-Dyson equation: 
\begin{equation}
    \begin{aligned}
        \Bigl\langle[\Omega_I^RE](x_0)\phi(x_1)\cdots \phi(x_n)\Bigr\rangle = -i\sum_{k=1}^n\delta(x_0-x_k)\Bigl\langle \Omega_I^R(x_0)\phi(x_1)\cdots\widehat{\phi(x_k)}\cdots \phi(x_n)\Bigr\rangle\ , 
    \end{aligned}
\end{equation}
where the hat symbol $\widehat{\phi(x_k)}$ indicates the omission of the field $\phi(x_k)$. This result confirms that the correlation functions of $\Omega_I^R E$ are UV finite, implying that $\Omega_I^R E$ itself is a renormalized operator. 

This line of reasoning can be extended to operators proportional to derivatives of $E$, for instance, $\Omega_I^R\partial_{\mu}E$, which we also classify as EOM operators. Their associated correlation functions remain UV finite, as shown below:
\begin{equation}
    \begin{aligned}
        &\Bigl\langle [\Omega_I^R\partial_{\mu}E](x_0)\phi(x_1)\cdots \phi(x_n)\Bigr\rangle\\
        =&-i\sum_{k=1}^n \frac{\partial}{\partial x_0^{\mu}}\delta(x_0-x_k)\Bigl\langle \Omega_I^R(x_0)\phi(x_1)\cdots \widehat{\phi(x_k)}\cdots \phi(x_n)\Bigr\rangle\ .\\
    \end{aligned}
\end{equation}

Given that the renormalization of EOM operators is fully determined by the renormalization of lower-dimensional operators, substituting certain operators in the basis with EOM operators allows part of the Z-matrix to be derived without additional computation.

As an illustration, consider the operator basis \eqref{operator basis 2}. The two dimension-4 operators can be combined to construct an EOM operator, which is then used to replace $\phi^{2}$. Since $\partial^{2}\phi$ is inherently a renormalized operator, the renormalized operator basis in this case is fully determined: 
\begin{equation}
    \begin{aligned}
        \Omega_{I}^{R}=\Bigl(E-\partial^{2}\phi,\ \partial^{2}\phi\Bigr)\ ,
    \end{aligned}
\end{equation}
where the renormalized operators are combined so that ${Z_{I}}^{J}=\delta_{I}^J+\mathcal{O}(g)$. By comparing the bare basis $\Omega_{I}^{0}$ and the renormalized basis $\Omega_{I}^{R}$, the Z-matrix can be directly obtained: 
\begin{equation}
    \begin{aligned}
        Z_I^{\ J}=
        \begin{pmatrix}
            Z_{\phi}^{\frac{1}{2}}\ \ \ \  &Z_{\phi}^{\frac{1}{2}}-Z_{\phi}^{-\frac{1}{2}}  \\
            0\ \ \ \  & Z_{\phi}^{-\frac{1}{2}}
        \end{pmatrix}\ .
    \end{aligned}
\end{equation}
The anomalous dimension matrix is proportional to $\gamma_{\phi}$: 
\begin{equation}\label{result-dim3}
    \begin{aligned}
        \gamma_I^{\ J}=
        \begin{pmatrix}
            -\gamma_{\phi}\ \ \ \  & -2\gamma_{\phi}  \\
            0\ \ \ \  & \gamma_{\phi}
        \end{pmatrix}\ , 
    \end{aligned}
\end{equation}
which is consistent to \eqref{gamma-dim-4}. 

It is more convenient to choose the bare operator basis as: 
\begin{equation}
    \begin{aligned}
        \Omega_{I}^{'0}=\Bigl(E_{0},\ \partial^2\phi_{0}\Bigr)\ .
    \end{aligned}
\end{equation}
Then the Z-matrix is diagonal: 
\begin{equation}
    \begin{aligned}
        {Z_{I}}^{J}=
        \begin{pmatrix}
            Z_{\phi}^{\frac{1}{2}}\ \ \ \  &0 \\
            0\ \ \ \  & Z_{\phi}^{-\frac{1}{2}}
        \end{pmatrix}\ .
    \end{aligned}
\end{equation}

In our computation, we choose both hard and soft operators to contain only monomial operators (e.g., those defined in \eqref{operator basis 2}). This choice simplifies the derivation of projection operators and OPE coefficients.
Once the anomalous dimension matrix is obtained, we perform a change of basis to distinguish between primary and descendant operators. The transition from the old basis to this new basis is governed by a constant matrix $T_I^{\ J}$, defined as follows:
\begin{equation}
    \begin{aligned}
        \Omega_{I}^{'0}={T_{I}}^{J}\Omega_{J}^{0},\ \Omega_{I}^{'R}={T_{I}}^{J}\Omega_{J}^{R}={(TZT^{-1})_{I}}^{J}\Omega_{J}^{'0}\ .
    \end{aligned}
\end{equation}
Under this transformation, the anomalous dimension matrix transforms as
\begin{equation}
    \begin{aligned}
        {\gamma_{I}^{\prime}}^{J}=&-\frac{{\partial(TZT^{-1})_{I}}^{K}}{\partial\ln\mu}{(TZ^{-1}T^{-1})_{K}}^{J}={(T\gamma T^{-1})_{I}}^{J}\ .
    \end{aligned}
\end{equation}
This change of basis significantly reduces operator mixing, resulting in a simpler matrix structure. Furthermore, entries involving only descendant operators can be directly determined from the anomalous dimensions of lower-dimensional operators, providing a consistency check for our results. 

The new basis enables us to focus only on the renormalization of primary operators. The primary operators with dimension $\Delta\leq 10$ were listed in the following table:
\begin{table}[H]
    \centering
    \begin{tabular}{|c |c|c|c|c|}
        \hline
        Length & 2 & 3 & 4 & 5  \\
        \hline
        primary & $\phi^{2}$ & $\phi^{3}$,\ $(\partial_{\mu\nu}\phi)^{2}\phi$ & $\phi^{4}$ & $\phi^{5}$  \\
        \hline
    \end{tabular}
    \caption{The primary operators of different length with dimension $\Delta\leq 10$ in $\phi^{3}$ theory. }
    \label{tab:primary-length}
\end{table}

%%%%%%%%%%%%%%%%%%%%%%%%%%%%%%%%%%%%%%%%%%%
\subsection{Dimension-6 operators\label{dim-6}}
%%%%%%%%%%%%%%%%%%%%%%%%%%%%%%%%%%%%%%%%%%%
We now turn our attention to dimension-6 operators. The monomial operator basis can be chosen as 
\begin{equation}\label{phi3-hard-basis-2}
    \begin{aligned}
        \Omega_{I}^{0}=\Bigl( \frac{g_{0}^{2}}{3!}\phi^{3}_{0},\ \frac{g_{0}}{2}\partial^{2}\phi^{2}_{0},\ \frac{g_{0}}{2}\phi_{0}\partial^{2}\phi_{0},\ (\partial^{2})^{2}\phi_{0} \Bigr)\ , 
    \end{aligned}
\end{equation}
and the soft basis is chosen as 
\begin{equation}\label{phi3-soft-basis-2}
    \begin{aligned}
        O_{\alpha}^{0}=\Bigl( I,\ \phi_{0},\ \frac{g_{0}}{2}\phi^{2}_{0},\ \partial^{2}\phi_{0} \Bigr)\ . 
    \end{aligned}
\end{equation}

The projection operators corresponding to \eqref{phi3-soft-basis-2} are 
\begin{equation}
    \begin{aligned}
        \mathcal{P}_{00}^{1}=1\ ,\ 
        \mathcal{P}_{01}^{2}=1\ ,\ 
        \mathcal{P}_{02}^{3}=\frac{1}{g_{0}}\ ,\ 
        \mathcal{P}_{01}^{4}=-\frac{1}{2D}\hat{\partial}_{1}^{2}\ .
    \end{aligned}
\end{equation}
Then the OPE coefficients take the following form 
\begin{equation}
    \begin{aligned}
        {C_{0I}}^{\alpha}(p)=\mathcal{P}_{0n}^{\alpha}H_{0I}(p;k_{1},\cdots,k_{n})\Big|_{k_{i}=0}\ . 
    \end{aligned}
\end{equation}

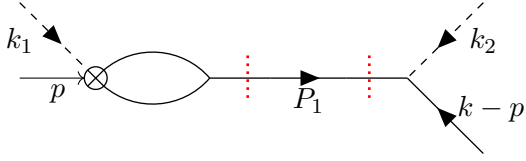
\begin{figure}
    \centering
    \begin{tikzpicture}
        \begin{feynhand}
            \vertex (a1) at (-1,1);
            \vertex[crossdot] (a0) at (0,0) {};
            \vertex (c1) at (-1,0);
            \draw [->] (c1) to (a0);
            \node at (-0.5,-0.2) {$p$};
            \propag[chasca] (a1) to (a0);
            \node at (-1,0.5) {$k_{1}$};
            \vertex (a2) at (1.5,0);
            \propag[plain] (a0) to [quarter left] (a2);
			\propag[plain] (a0) to [quarter right] (a2);
            \vertex (a3) at (2.3,0);
            \vertex (a4) at (3.3,0);
            \vertex (a5) at (4.1,0);
            \propag[plain] (a2) to (a3);
            %\node at (1.8,-0.3) {$P_{1}$};
            \propag[fer] (a3) to (a4);
            \node at (2.8,-0.3) {$P_{1}$};
            \propag[plain] (a4) to (a5);
            %\node at (3.7,-0.3) {$P_{1}$};
            \vertex (a6) at (5.1,-1);
            \propag[fer] (a6) to (a5);
            \node at (5.2,-0.4) {$k-p$};
            \vertex (a7) at (5.1,1);
            \propag[chasca] (a7) to (a5);
            \node at (5.1,0.5) {$k_{2}$};
            \vertex (b1) at (2.0,0.3);
			\vertex (b2) at (2.0,-0.3);
			\vertex (b3) at (3.6,0.3);
			\vertex (b4) at (3.6,-0.3);
			\propag[gho, red] (b1) to (b2);
			\propag[gho, red] (b3) to (b4);
        \end{feynhand}
    \end{tikzpicture}
    \caption{One-loop 1PR diagram contributing to the correlation function $\langle\Omega_{1}^{0}\phi\phi\phi\rangle$, where $P_{1}=k_{1}+p$. The red dot lines split the diagram into three parts corresponding to the three function on the RHS of \eqref{example 1PR one-loop} respectively. }
    \label{fig:one-loop 1PR}
\end{figure}

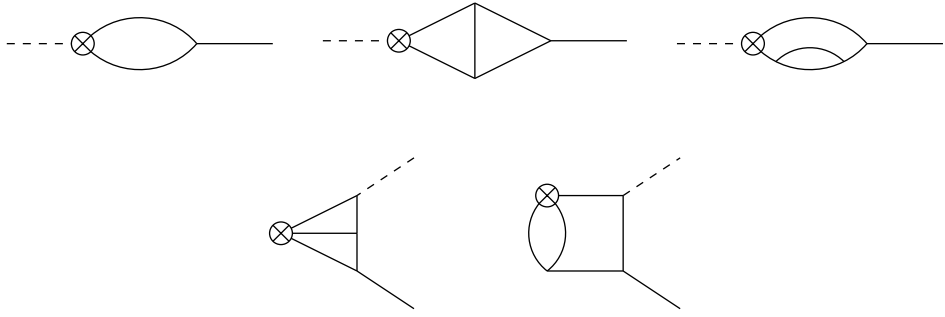
\begin{figure}
	\centering
	\subfloat
	{
		\begin{tikzpicture}
			\begin{feynhand}
				\vertex (a1) at (-1,0);
				\vertex[crossdot] (a0) at (0,0) {};
				\propag[sca] (a1) to (a0);
				\vertex (a2) at (1.5,0);
				\propag[plain] (a0) to [quarter left] (a2);
				\propag[plain] (a0) to [quarter right] (a2);
				\vertex (a3) at (2.5,0);
				\propag[plain] (a2) to (a3);
			\end{feynhand}
		\end{tikzpicture}
	}
	\quad
	\vspace{0.5cm}
	\subfloat
	{
		\begin{tikzpicture}
			\begin{feynhand}
				\vertex (a1) at (-1,0);
				\vertex[crossdot] (a0) at (0,0) {};
				\propag[sca] (a1) to (a0);
				\vertex (a2) at (1,0.5);
				\vertex (a3) at (1,-0.5);
				\propag[plain] (a2) to (a0);
				\propag[plain] (a3) to (a0);
				\propag[plain] (a2) to (a3);
				\vertex (a4) at (2,0);
				\propag[plain] (a2) to (a4);
				\propag[plain] (a3) to (a4);
				\vertex (a5) at (3,0);
				\propag[plain] (a4) to (a5);
			\end{feynhand}
		\end{tikzpicture}
	}
	\quad
	\vspace{0.5cm}
	\subfloat
	{
		\begin{tikzpicture}
			\begin{feynhand}
				\vertex (a1) at (-1,0);
				\vertex[crossdot] (a0) at (0,0) {};
				\propag[sca] (a1) to (a0);
				\vertex (a2) at (1.5,0);
				\propag[plain] (a0) to [quarter left] (a2);
				\propag[plain] (a0) to [quarter right] (a2);
				\vertex (a3) at (2.5,0);
				\propag[plain] (a2) to (a3);
				\vertex (a4) at (0.3,-0.24);
				\vertex (a5) at (1.2,-0.24);
				\propag[plain] (a4) to [quarter left] (a5);
			\end{feynhand}
		\end{tikzpicture}
	}\\
	\subfloat
	{
		\begin{tikzpicture}
			\begin{feynhand}
				\vertex[crossdot] (a0) at (0,0) {};
				\vertex (a2) at (1,0.5);
				\vertex (a3) at (1,-0.5);
				\propag[plain] (a2) to (a0);
				\propag[plain] (a3) to (a0);
				\vertex (a4) at (1,0);
				\propag[plain] (a2) to (a4);
				\propag[plain] (a3) to (a4);
				\propag[plain] (a0) to (a4);
				\vertex (a1) at (1.75,1);
				\vertex (a5) at (1.75,-1);
				\propag[sca] (a1) to (a2);
				\propag[plain] (a5) to (a3);
			\end{feynhand}
		\end{tikzpicture}
	}
	\quad\quad\quad
	\subfloat
	{
		\begin{tikzpicture}
			\begin{feynhand}
				\vertex[crossdot] (a0) at (0,0) {};
				\vertex (a1) at (1,0);
				\vertex (a2) at (1,-1);
				\vertex (a3) at (0,-1);
				\propag[plain] (a0) to (a1);
				\propag[plain] (a1) to (a2);
				\propag[plain] (a2) to (a3);
				\propag[plain] (a0) to [quarter left] (a3);
				\propag[plain] (a0) to [quarter right] (a3);
				\vertex (a4) at (1.75,0.5);
				\vertex (a5) at (1.75,-1.5);
				\propag[sca] (a1) to (a4);
				\propag[plain] (a2) to (a5);
			\end{feynhand}
		\end{tikzpicture}
	}
	\caption{The one-loop and two-loop diagrams of $G(p,-P_{1})$. There is no tree-level diagram. }
	\label{fig:1PI-1}
\end{figure}

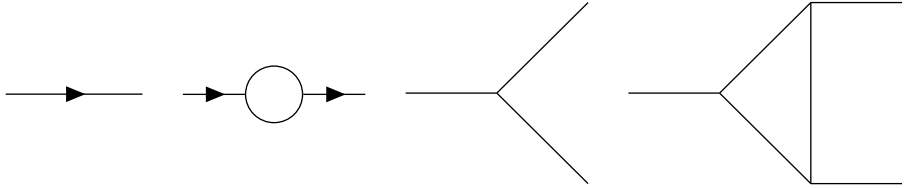
\begin{figure}
    \centering
    \subfloat{
    \raisebox{31pt}{\begin{tikzpicture}[scale=1.2]
        \begin{feynhand}
            \vertex (a1) at (0,0);
            \vertex (a2) at (1.5,0);
            \propag[fer] (a1) to (a2);
        \end{feynhand}
    \end{tikzpicture}}}
    \quad
    \subfloat{
    \raisebox{23pt}{\begin{tikzpicture}[scale=1.2]
        \begin{feynhand}
            \vertex (a1) at (0,0);
            \vertex[ringblob] (a3) at (1,0) {};
            \vertex (a2) at (2,0);
            \propag[fer] (a1) to (a3);
            \propag[fer] (a3) to (a2);
        \end{feynhand}
    \end{tikzpicture}}}
    \quad
    \subfloat{
    \raisebox{0pt}{\begin{tikzpicture}[scale=1.2]
        \begin{feynhand}
            \vertex (a1) at (-1,0);
            \vertex (a0) at (0,0);
            \vertex (a2) at (1,1);
            \vertex (a3) at (1,-1);
            \propag[plain] (a1) to (a0);
            \propag[plain] (a2) to (a0);
            \propag[plain] (a3) to (a0);
        \end{feynhand}
    \end{tikzpicture}}}
    \quad
    \subfloat{
    \raisebox{0pt}{\begin{tikzpicture}[scale=1.2]
        \begin{feynhand}
            \vertex (a1) at (-1,0);
            \vertex (a0) at (0,0);
            \vertex (a2) at (1,1);
            \vertex (a3) at (1,-1);
            \propag[plain] (a1) to (a0);
            \propag[plain] (a2) to (a0);
            \propag[plain] (a3) to (a0);
            \propag[plain] (a2) to (a3);
            \vertex (a4) at (2,1);
            \vertex (a5) at (2,-1);
            \propag[plain] (a4) to (a2);
            \propag[plain] (a5) to (a3);
        \end{feynhand}
    \end{tikzpicture}}}
    \caption{tree-level and one-loop exact-propagators and three-point correlation function of fundamental fields.}
    \label{fig:1PI-2}
\end{figure}

As discussed in Section \ref{section:1PR}, 1PR diagrams contribute to the hard correlation functions. For instance, the four-point correlation function $\langle\Omega_{1}^{0}\phi\phi\phi\rangle$ has the following contribution, 
\begin{equation}\label{example 1PR one-loop}
    \begin{aligned}
        H^{1PR}_{10}(p,k-p;k_{1},k_{2})=G(p,-P_{1})\frac{1}{(P_{1})^{2}}G(P_{1},k-p;k_{2})+\Bigl( k_{1}\leftrightarrow k_{2}\Bigr)\ . 
    \end{aligned}
\end{equation}
The contribution of the one-loop diagram in Figure \ref{fig:one-loop 1PR} can be determined by separately computing each of its three 1PI parts. For 1PR diagrams of higher loops, we can also obtain the contributing 1PR diagrams by connecting the three corresponding 1PI components displayed in Figures \ref{fig:1PI-1} and \ref{fig:1PI-2} (up to two loops). 

Three independent descendant operators can be constructed from the linear combination of dimension-6 operators, and the new operator basis is chosen as
\begin{equation}
   \begin{aligned}
       \Omega_{I}^{\Delta=6;0}=\Bigl(\frac{g_{0}^{2}}{3!}\phi_{0}^{3},\ g_{0}\phi_{0}E_{0},\ \partial^{2}E_{0},\ (\partial^{2})^{2}\phi_{0}\Bigr)\ . 
   \end{aligned}
\end{equation}
The corresponding Z-matrix takes the form: 
\begin{equation}
    \begin{aligned}
        {Z_{I}}^{J}=
        \begin{pmatrix}
            {Z_{1}}^{1}\ \ \ \  & {Z_{1}}^{2}\ \ \ \  & {Z_{1}}^{3}\ \ \ \  & {Z_{1}}^{4} \\
            0\ \ \ \  & \tilde{Z}_{g}^{-1}\ \ \ \  & 0\ \ \ \  & 0 \\
            0\ \ \ \  & 0\ \ \ \  & Z_{\phi}^{\frac{1}{2}}\ \ \ \  & 0 \\
            0\ \ \ \  & 0\ \ \ \  & 0\ \ \ \  & Z_{\phi}^{-\frac{1}{2}} \\
        \end{pmatrix}\ , 
    \end{aligned}
\end{equation}
where $\tilde{Z}_{g}$ is defined through
\begin{equation}
    g_0=\tilde{Z}_gg\tilde{\mu}^{\epsilon}\ .
\end{equation}

Since the matrix is upper-triangular, the anomalous dimensions can be completely determined by the diagonal entries. The anomalous dimension matrix is given by 
\begin{equation}
    \begin{aligned}
        {\gamma_{I}}^{J}=
        \begin{pmatrix}
            -\frac{3g^{2}}{4}-\frac{125g^{4}}{48}\ \ \ \  & -\frac{g^{2}}{6}-\frac{13g^{4}}{108}\ \ \ \  & -\frac{g^{4}}{36}\ \ \ \  & \frac{5g^{4}}{144} \\
            0\ \ \ \  & -\beta(g)/g\ \ \ \  & 0\ \ \ \  & 0 \\
            0\ \ \ \  & 0\ \ \ \  & -\gamma_{\phi}\ \ \ \  & 0 \\
            0\ \ \ \  & 0\ \ \ \  & 0\ \ \ \  & \gamma_{\phi}
        \end{pmatrix}\ . 
    \end{aligned}
\end{equation}

The presence of the primary operator $\phi^{3}$ in the Lagrangian directly links its renormalization to the beta function.
The connection can be expressed as
\begin{equation}\label{phi3-dim6-beta}
    {\gamma_{1}}^{1}=g^{2}\frac{\partial}{\partial g}\Bigl(\frac{\beta(g)}{g^{2}}\Bigr)
\end{equation}
This follows a general pattern in quantum field theories: the renormalization of marginal operators can be derived from the corresponding beta functions. The detailed derivation is provided in Appendix \ref{appendixA}.

%%%%%%%%%%%%%%%%%%%%%%%%%%%%%%%%%%%%%%%%%%%
\subsection{Dimension-8 and dimension 10 operators\label{dim-8-10}}
%%%%%%%%%%%%%%%%%%%%%%%%%%%%%%%%%%%%%%%%%%%

%%%%%%%%%%%%%%%%%%%%%%%%%%%%%%%%%%%%%%%%%%%
\paragraph{Dimension-8 operators}
%%%%%%%%%%%%%%%%%%%%%%%%%%%%%%%%%%%%%%%%%%%
We choose the basis of dimension-8 bare operators as
\begin{equation}
    \begin{aligned}
        \Omega_I^{\Delta=8;0}=\Bigl(\frac{g_{0}^{3}}{4!}\phi_{0}^{4},\ g_{0}\phi_{0}\partial^{2}E_{0}\Bigr)\cup g_{0}E_{0}\Omega_{I}^{\Delta=4;0}\cup \partial^{2}\Omega_{I}^{\Delta=6;0}\ , 
    \end{aligned}
\end{equation}
where $\Omega^{\Delta=4;0}=\Bigl(E_{0},\ \partial^2\phi_{0}\Bigr)$. The soft basis is chosen as 
\begin{equation}
    \begin{aligned}
        O_{\alpha}^{0}=\Bigl( I,\ \phi_{0},\ \partial_{\mu}\phi_{0},\ \frac{g_{0}}{2}\phi^{2}_{0},\ \partial^{2}\phi_{0},\ \frac{g_{0}^{2}}{3!}\phi^{3}_{0},\ \frac{g_{0}}{2}\partial^{2}\phi^{2}_{0},\ \frac{g_{0}}{2}\phi_{0}\partial^{2}\phi_{0} \Bigr)\ . 
    \end{aligned}
\end{equation}
Here we choose a tensor operator in soft basis, as we have proved previously, our algorithm applies to tensor operators. 

The only primary operator is $\Omega_1$, for which
\begin{equation}
    \begin{aligned}
        {\gamma_{1}}^{J}=\Bigl(-\frac{29g^{2}}{12}-\frac{4061g^{4}}{432},\ \frac{g^{4}}{144},\ -\frac{g^{2}}{6}-\frac{31g^{4}}{144},\ \frac{g^{2}}{6}+\frac{31g^{4}}{144},\ -\frac{5g^{4}}{144},\ 0,\ 0,\ 0\Bigr)\ . 
    \end{aligned}
\end{equation}
All other entries of ${\gamma_{I}}^{J}$ can be determined directly from the anomalous dimension matrices of lower dimensional operators. 

%%%%%%%%%%%%%%%%%%%%%%%%%%%%%%%%%%%%%%%%%%%
\paragraph{Dimension-10 operators}
%%%%%%%%%%%%%%%%%%%%%%%%%%%%%%%%%%%%%%%%%%%
We choose the basis of dimension-10 bare operators as 
\begin{equation}
    \begin{aligned}
        \Omega_{I}^{\Delta=10;0}=\Bigl(\frac{g_{0}^{4}}{5!}\phi_{0}^{5},\ 
        \frac{g_{0}^{2}}{2}\phi_{0}\partial_{\mu\nu}\phi_{0}\partial^{\mu\nu}\phi_{0},\ g_{0}\phi_{0}(\partial^{2})^{2}E_{0},\ g_{0}\partial^{2}\phi_{0}\partial^{2}E_{0}\Bigr)\cup g_{0}E_{0} \Omega_{I}^{\Delta=6;0}\cup \partial^{2} \Omega_{I}^{\Delta=8;0}\ . 
    \end{aligned}
\end{equation}
And we choose the following soft basis, 
\begin{equation}
    \begin{aligned}
        O_{\alpha}^{0}=\Bigl( I,\ \phi_{0},&\ \partial_{\mu}\phi_{0},\ \frac{g_{0}}{2}\phi^{2}_{0},\ \partial^{2}\phi_{0},\ \frac{g_{0}^{2}}{3!}\phi^{3}_{0},\ \frac{g_{0}}{2}\partial^{2}\phi^{2}_{0},\ \frac{g_{0}}{2}\phi_{0}\partial^{2}\phi_{0},\ \frac{g_{0}^{3}}{4!}\phi^{4},\ \frac{g_{0}^{2}}{3!}\partial^{2}(\phi^{3}),\ \\
        &\frac{g_{0}^{2}}{3!}\partial^{2}\phi\phi^{2},\ \frac{g_{0}}{2}(\partial^{2})^{2}\phi^{2},\ \frac{g_{0}}{2}(\partial^{2})^{2}\phi\phi,\ \frac{g_{0}}{2}\partial^{2}\phi\partial^{2}\phi,\ \frac{g_{0}}{2}\partial^{2}\partial_{\mu}\phi\partial^{\mu}\phi,\ (\partial^{2})^{3}\phi\Bigr)\ . 
    \end{aligned}
\end{equation}

The first two operators are primaries, for which:
\begin{equation}
    \begin{aligned}
        {\gamma_{1}}^{J}=\Bigl(-\frac{59g^{2}}{12}-\frac{9515g^{4}}{432},&\ 0,\ 0,\ -\frac{g^{4}}{144},\ -\frac{g^{2}}{6}-\frac{67g^{4}}{216},\ \\
        &0,\ \frac{g^{4}}{144},\ 0,\ -\frac{5g^{4}}{144},\ 0,\ 0,\ 0,\ 0,\ 0,\ 0,\ 0\Bigr)\ , 
    \end{aligned}
\end{equation}
and
\begin{equation}
	\begin{aligned}
		{\gamma_{2}}^{J}=\Bigl(&-\frac{153g^{2}}{4}-\frac{393037g^{4}}{2400},\ \frac{11g^{2}}{60}-\frac{1283g^{4}}{6750},\ -\frac{g^{2}}{15}-\frac{4043g^{4}}{108000},\ \frac{11g^{2}}{40}+\frac{20687g^{4}}{72000},\   \\
		&-\frac{5g^{2}}{4}-\frac{335g^{4}}{144},\ -\frac{37g^{2}}{120}-\frac{102389g^{4}}{216000},\ \frac{3g^{2}}{10}+\frac{19213g^{4}}{24000},\ -\frac{19g^{2}}{30}-\frac{118321g^{4}}{108000},\   \\
		&-\frac{13g^{2}}{4}-\frac{162673g^{4}}{14400},\ \frac{g^{2}}{20}+\frac{1139g^{4}}{72000},\ -\frac{41g^{2}}{120}-\frac{130337g^{4}}{216000},\ \frac{47g^{2}}{120}+\frac{138679g^{4}}{216000},\   \\
		&-\frac{g^{2}}{10}-\frac{613g^{4}}{2000},\ -\frac{g^{2}}{40}-\frac{61g^{4}}{6000},\ -\frac{31g^{4}}{10800},\ \frac{53g^{4}}{14400}\Bigr)\ . 
	\end{aligned}
\end{equation}

%%%%%%%%%%%%%%%%%%%%%%%%%%%%%%%%%%%%%%%%%%%
\section{\texorpdfstring{Renormalization of operators in $\phi^{4}$ model}{Renormalization of operators in φ4 model}}\label{re-phi3phi4}
%%%%%%%%%%%%%%%%%%%%%%%%%%%%%%%%%%%%%%%%%%%
The renormalization of operators within the $\phi^4$ model can be investigated using the methodology outlined in previous sections. This section presents the anomalous dimensions of operators ranging from dimension-1 to dimension-5, calculated up to five-loop order. In addition to IBP reduction, we employed graphical functions~\cite{Schnetz:2013hqa,Golz:2015rea,Borinsky:2021gkd,Schnetz:2024qqt}, which demonstrates superior efficiency at higher loop orders, to evaluate several two-point integrals encountered during the computation. With this approach, the gamma function in $\phi^4$ theory has been computed through eight loops in the minimal subtraction scheme, while the beta function, the mass anomalous dimension, and the self-energy have been obtained through seven loops~\cite{Schnetz:2022nsc}. 

%%%%%%%%%%%%%%%%%%%%%%%%%%%%%%%%%%%%%%%%%%%
\paragraph{Field anomalous dimension and the beta function}
%%%%%%%%%%%%%%%%%%%%%%%%%%%%%%%%%%%%%%%%%%%
The anomalous dimension of the fundamental field $\phi$, as determined from the two-point correlation function, is given by:
\begin{equation}
    \begin{aligned}
        \gamma_{\phi}=0g+\frac{g^{2}}{12}-\frac{g^{3}}{16}+\frac{65 g^{4}}{192}+\left(\frac{3 \zeta_{3}}{16}-\frac{\pi^{4}}{180}-\frac{3709}{2304}\right) g^{5}\ . 
    \end{aligned}
\end{equation}

The beta function is again derived by employing the OPE of two fundamental fields:
\begin{equation}
    \phi(k-p)\phi(p)\sim \frac{1}{2}C(p)\phi_R^2(k)\ .
\end{equation}
In the current case, the composite operator $\phi^2$ appears as the soft operator. Consequently, its $Z$-factor appears in the ratio between renormalized and bare OPE coefficients:
\begin{equation}
    C(p)=\frac{Z_{\phi}}{Z_{\phi^2}}C_0(g_0;p)\ .
\end{equation}
The anomalous dimension of $\phi^2$ will be given in \eqref{gamma-phi2}. Using this result, the beta function can be obtained as:
\begin{equation}
    \begin{aligned}
        \frac{\partial g}{\partial \ln\mu}=-2 \epsilon g&+3 g^2-\frac{17 g^{3}}{3}+\left(12 \zeta_{3}+\frac{145}{8}\right) g^{4}+\left(-78 \zeta_{3}-120 \zeta_{5}+\frac{\pi^{4}}{5}-\frac{3499}{48}\right) g^5\\
        &+\left(45 \zeta_{3}^2+\frac{7965 \zeta_{3}}{16}+987 \zeta_{5}+1323 \zeta_{7}-\frac{5 \pi^{6}}{14}-\frac{1189 \pi^{4}}{720}+\frac{764621}{2304}\right) g^{6}\ , 
    \end{aligned}
\end{equation}

%%%%%%%%%%%%%%%%%%%%%%%%%%%%%%%%%%%%%%%%%%%
\paragraph{Dimension-2 operators}
%%%%%%%%%%%%%%%%%%%%%%%%%%%%%%%%%%%%%%%%%%%
The dimension-2 operator in $\phi^{4}$ model is
\begin{equation}
    \begin{aligned}
        \Omega_{I}^{0}=\frac{1}{2}\phi^{2}_{0}\ . 
    \end{aligned}
\end{equation}
The soft basis is chosen as 
\begin{equation}
    \begin{aligned}
        O_{\alpha}^{0}=\phi_{0}\ . 
    \end{aligned}
\end{equation}
We consider the OPE of $\Omega_{I}$ and $\phi$: 
\begin{equation}
    \begin{aligned}
        \frac{1}{2}\phi^{2}(p)\phi(k-p)\sim C(p)\phi+\cdots
    \end{aligned}
\end{equation}
And the five-loop anomalous dimension of $\phi^{2}$ is 
\begin{equation}
    \begin{aligned}
        \gamma_{\phi^{2}}=g-&\frac{5 g^2}{6}+\frac{7 g^3}{2}-\left(\frac{3 \zeta_{3}}{2}+\frac{\pi^{4}}{30}+\frac{477}{32}\right) g^{4}  \\
        &-\left(9 \zeta_{3}^2-\frac{1519 \zeta_{3}}{48}-\zeta_{5}-\frac{13 \pi^{4}}{72}-\frac{5 \pi^{6}}{126}-\frac{158849}{2304}\right) g^{5}\ . 
    \end{aligned}\label{gamma-phi2}
\end{equation} 

%%%%%%%%%%%%%%%%%%%%%%%%%%%%%%%%%%%%%%%%%%%
\paragraph{Dimension-3 operators}
%%%%%%%%%%%%%%%%%%%%%%%%%%%%%%%%%%%%%%%%%%%
We choose the dimension-3 operator basis as 
\begin{equation}
    \begin{aligned}
        \Omega_{I}^{0}=\Bigl(E_{0},\ \partial^2\phi_{0}\Bigr)\ . 
    \end{aligned}
\end{equation}
And the soft basis can be chosen as 
\begin{equation}
    \begin{aligned}
        O_{\alpha}^{0}=\Bigl( I,\ \frac{1}{2}\phi^{2}_{0} \Bigr)\ . 
    \end{aligned}
\end{equation}
Then the Z-matrix is diagonal: 
\begin{equation}
	\begin{aligned}
		{Z_{I}}^{J}=
		\begin{pmatrix}
			Z_{\phi}^{\frac{1}{2}}\ \ \ \  &0 \\
			0\ \ \ \  & Z_{\phi}^{-\frac{1}{2}}
		\end{pmatrix}\ .
	\end{aligned}
\end{equation}
And the anomalous dimensions are $-\gamma_{\phi}$ and $\gamma_{\phi}$ respectively. 

%%%%%%%%%%%%%%%%%%%%%%%%%%%%%%%%%%%%%%%%%%%
\paragraph{Dimension-4 operators}
%%%%%%%%%%%%%%%%%%%%%%%%%%%%%%%%%%%%%%%%%%%
The bare operator basis can be chosen as 
\begin{equation}
    \begin{aligned}
        \Omega_{I}^{0}=\Bigl(\frac{g_{0}}{4!}\phi_{0}^{4},\ \frac{1}{2}\phi_{0}E_{0},\ \frac{1}{2}\partial^{2}\phi_{0}^{2}\Bigr)\ .
    \end{aligned}
\end{equation}
We choose the soft basis as 
\begin{equation}
    \begin{aligned}
        O_{\alpha}^{0}=\Bigl( \phi_{0},\ \frac{g_{0}}{3!}\phi^{3}_{0},\ \partial^{2}\phi_{0} \Bigr)\ . 
    \end{aligned}
\end{equation}
The Z-matrix takes the following form: 
\begin{equation}
    \begin{aligned}
        {Z_{I}}^{J}=
        \begin{pmatrix}
            {Z_{1}}^{1}\ \ \ \  & {Z_{1}}^{2}\ \ \ \  & {Z_{1}}^{3} \\
            0\ \ \ \  & 1\ \ \ \  & 0 \\
            0\ \ \ \  & 0\ \ \ \  & Z_{\phi^2}\\
        \end{pmatrix}\ .
    \end{aligned}
\end{equation}
And the ${\gamma_{I}}^{J}$ is 
\begin{equation}
{\gamma_{I}}^{J}=
    \begin{pmatrix}
        g\frac{\partial}{\partial g}\frac{\beta(g)}{g}\ \ \ \  & \frac{g^{2}}{3}-\frac{3g^{3}}{8}+\frac{65g^{4}}{24}\ \ \ \  & -\frac{g^{3}}{6}+\frac{11g^{4}}{9} \\
        0\ \ \ \  & 0\ \ \ \  & 0 \\
        0\ \ \ \  & 0\ \ \ \  & \gamma_{\phi^{2}} \\
    \end{pmatrix}\ . 
\end{equation}

%%%%%%%%%%%%%%%%%%%%%%%%%%%%%%%%%%%%%%%%%%%
\paragraph{Dimension-5 operators}
%%%%%%%%%%%%%%%%%%%%%%%%%%%%%%%%%%%%%%%%%%%
The bare operator basis is chosen as
\begin{equation}
    \begin{aligned}
        \Omega_{I}^{0}=\Bigl(\frac{g_{0}^{2}}{5!}\phi_{0}^{5},\ \frac{g_{0}}{2}\phi_{0}^{2}E_{0},\ \partial^{2}E_{0},\ (\partial^{2})^{2}\phi_{0}\Bigr)\ . 
    \end{aligned}
\end{equation}
The soft basis is chosen as 
\begin{equation}
    \begin{aligned}
        O_{\alpha}^{0}=\Bigl( I,\ \frac{1}{2}\phi^{2}_{0},\ \frac{g_{0}}{4!}\phi^{4}_{0},\ \frac{1}{2}\partial^{2}\phi^{2}_{0},\ \frac{1}{2}\phi_{0}\partial^{2}\phi_{0} \Bigr)\ . 
    \end{aligned}
\end{equation}
The Z-matrix takes the form: 
\begin{equation}
    \begin{aligned}
        {Z_{I}}^{J}=
        \begin{pmatrix}
            {Z_{1}}^{1}\ \ \ \  & {Z_{1}}^{2}\ \ \ \  & {Z_{1}}^{3}\ \ \ \  & {Z_{1}}^{4} \\
            0\ \ \ \  & \tilde{Z}_{g}^{-1}Z_{\phi}^{\frac{1}{2}}Z_{\phi^2}\ \ \ \  & 0\ \ \ \  & 0 \\
            0\ \ \ \  & 0\ \ \ \  & Z_{\phi}^{\frac{1}{2}}\ \ \ \  & 0 \\
            0\ \ \ \  & 0\ \ \ \  & 0\ \ \ \  & Z_{\phi}^{-\frac{1}{2}} \\
        \end{pmatrix}\ , 
    \end{aligned}
\end{equation}
where $\tilde{Z}_{g}$ is a renormalization constant defined in Appendix \ref{appendixA}. Using the definition of beta function 
\begin{equation}
    \begin{aligned}
        \frac{\partial g}{\partial\ln \mu}=\frac{-2\epsilon g}{1+g\frac{\partial \ln \tilde{Z}_{g}}{\partial g}}\ , 
    \end{aligned}
\end{equation}
We find 
\begin{equation}
    \begin{aligned}
        \frac{\partial \ln \tilde{Z}_g}{\partial\ln \mu}=-2\epsilon-\frac{\partial \ln g}{\partial\ln \mu}=-\frac{\beta(g)}{g}\ , 
    \end{aligned}
\end{equation}
so that
\begin{equation}
    \begin{aligned}
        {\gamma_{2}}^{2}=-\frac{\beta(g)}{g}-\gamma_{\phi}+\gamma_{\phi^2}\ . 
    \end{aligned}
\end{equation}

\begin{equation}
    \begin{aligned}
        {\gamma_{1}}^{J}=\Bigl(&4g-\frac{319g^{2}}{12}+g^{3}(\frac{8605}{48}+96\zeta_3)+g^{4}(\frac{-257903}{192}+\frac{34\pi^{4}}{15}-1029\zeta_3-1560\zeta_5),\ \\
        &-\frac{g^{2}}{6}+\frac{5g^{3}}{12}-\frac{169g^{4}}{48},\ \frac{g^{3}}{6}-\frac{265g^{4}}{144},\ -\frac{g^{3}}{6}+\frac{133g^{4}}{72}\Bigr)\ . 
    \end{aligned}
\end{equation}

%%%%%%%%%%%%%%%%%%%%%%%%%%%%%%%%%%%%%%%%%%%
\section{Checks}\label{checks}
%%%%%%%%%%%%%%%%%%%%%%%%%%%%%%%%%%%%%%%%%%%

%%%%%%%%%%%%%%%%%%%%%%%%%%%%%%%%%%%%%%%%%%%
\paragraph{\texorpdfstring{$\phi^{3}$ model}{φ3 model}}
%%%%%%%%%%%%%%%%%%%%%%%%%%%%%%%%%%%%%%%%%%%
Our results for the field anomalous dimension and the beta function agree through two loops with Refs.~\cite{Gracey:2015tta,Pang:2016xno,Kompaniets:2021hwg,Borinsky:2021jdb}, after matching the coupling conventions. For the dimension-4 operator, the comparison with the mass anomalous dimension is performed after removing the explicit coupling factor from $g\phi^2/2$ and including the corresponding $\beta(g)/g$
contribution. 

The dimension-6 marginal-operator result satisfies the beta-function relation~\eqref{phi3-dim6-beta}. The descendant and EOM operators are consistently reproduced from the renormalization of lower-dimensional operators. 

To the best of our knowledge, the two-loop results for the dimension-8 and dimension-10 operators listed in Table~\ref{tab:checks-phi3} have not been reported previously for the one-component six-dimensional $\phi^3$ theory. The higher-loop results of Ref.~\cite{Huang:2025rdy} concern tensor operators in an $O(N)$ cubic theory with both $\phi_i$ and $\sigma$ fields and therefore do not provide a direct comparison with the present one-component theory. 

\begin{table}[t]
	\centering
	\renewcommand{\arraystretch}{1.45}
	\setlength{\tabcolsep}{6pt}
	\begin{tabular}{
			>{\raggedright\arraybackslash}m{0.38\textwidth}
			>{\raggedright\arraybackslash}m{0.56\textwidth}}
		\hline
		\textbf{Operator} & \textbf{Comparison or check} \\
		\hline
		
		$\gamma_{\phi}$ and $\beta(g)$
		&
		Agreement through two loops with
		Refs.~\cite{Gracey:2015tta,Pang:2016xno,
			Kompaniets:2021hwg,Borinsky:2021jdb}.
		\\ \hline
		
		$\displaystyle \frac{g}{2}\phi^2$
		&
		Agreement through two loops with
		Refs.~\cite{Kompaniets:2021hwg,Borinsky:2021jdb},
		after matching normalization conventions.
		\\ \hline
		
		$\displaystyle \frac{g^2}{3!}\phi^3$
		&
		Satisfies the relation between anomalous dimension and beta-function
		in Eq.~\eqref{phi3-dim6-beta}.
		\\ \hline
		
		$\displaystyle \frac{g^3}{4!}\phi^4$
		&
		New two-loop correction in the one-component
		$\phi^3$ theory, to the best of our knowledge.
		\\ \hline
		
		$\displaystyle
		\frac{g^4}{5!}\phi^5,\ 
		\frac{g^2}{2}\phi\,\partial_{\mu\nu}\phi\,
		\partial^{\mu\nu}\phi$
		&
		New two-loop corrections in the one-component
		$\phi^3$ theory, to the best of our knowledge.
		For comparison, the higher-loop results of
		Ref.~\cite{Huang:2025rdy} concern a different
		$O(N)$ cubic theory.
		\\ \hline
		
		Descendant and EOM operators
		&
		Consistent with the lower-dimensional renormalization
		relations derived in Section~\ref{re-descedant}.
		\\ \hline
		
	\end{tabular}
	\caption{Comparisons and consistency checks for the
		one-component six-dimensional $\phi^3$ theory.}
	\label{tab:checks-phi3}
\end{table}

%%%%%%%%%%%%%%%%%%%%%%%%%%%%%%%%%%%%%%%%%%%
\paragraph{\texorpdfstring{$\phi^{4}$ model}{φ4 model}}
%%%%%%%%%%%%%%%%%%%%%%%%%%%%%%%%%%%%%%%%%%%
Using the OPE method, we obtain the beta function $\beta(g)$ and the field anomalous dimension $\gamma_{\phi}$ through five loops. After matching coupling conventions, these results agree through five loops with Refs.~\cite{Kleinert:1991rg,Kompaniets:2017yct,Henriksson:2025vyi}. In particular, Ref.~\cite{Kompaniets:2017yct} gives the renormalization-group functions of the $O(n)$-symmetric $\phi^{4}$ theory through six loops. Setting $n=1$ gives the one-component theory considered here; we compare our results with this reference through five loops, the highest order obtained in our calculation. The anomalous dimension of $\frac{1}{2}\phi^{2}$ agrees through five loops with Refs.~\cite{Kleinert:1991rg,Kompaniets:2017yct,Henriksson:2025vyi}. The dimension-3 operators agree with Ref.~\cite{Henriksson:2025vyi} through five loops. 

The dimension-4 marginal operator $\frac{g}{4!}\phi^{4}$ satisfies the general relation~\eqref{gamma-beta}; at the fixed point $g^{*}$ where $\beta(g^{*})=0$, this reduces to $\gamma=\frac{\partial \beta(g)}{\partial g}\big|_{g=g^{*}}$, in agreement with Ref.~\cite{Henriksson:2025vyi}. The dimension-5 operator $\frac{g^{2}}{5!}\phi^{5}$ agrees with Ref.~\cite{Henriksson:2025vyi} through four loops. For the dimension-5 operator, the comparison is performed after removing the explicit coupling factor from the operator definition and including the corresponding $2\beta(g)/g$ contribution. The descendant and EOM operators are consistently reproduced from the renormalization of lower-dimensional operators.

The dimension-2 to dimension-5 results listed in Table~\ref{tab:checks-phi4} reproduce the corresponding results of Ref.~\cite{Henriksson:2025vyi} and, at lower orders, of the earlier literature; our results do not constitute new results for the four-dimensional $\phi^{4}$ theory. Rather, our results provide an independent confirmation of the present framework in a theory whose operator content differs from that of the $\phi^{3}$ model. 

\begin{table}[t]
	\centering
	\renewcommand{\arraystretch}{1.45}
	\setlength{\tabcolsep}{6pt}
	\begin{tabular}{
			>{\raggedright\arraybackslash}m{0.38\textwidth}
			>{\raggedright\arraybackslash}m{0.56\textwidth}}
		\hline
		\textbf{Operator} & \textbf{Comparison or check} \\
		\hline
		
		$\gamma_{\phi}$ and $\beta(g)$
		&
		Agreement through five loops with
		Refs.~\cite{Kleinert:1991rg,Kompaniets:2017yct,Henriksson:2025vyi}.
		\\ \hline
		
		$\displaystyle \frac{1}{2}\phi^{2}$
		&
		Agreement through five loops with
		Ref.~\cite{Henriksson:2025vyi},
		after matching normalization conventions.
		\\ \hline
		
		$\displaystyle E,\ \partial^{2}\phi$
		&
		Agreement through five loops with
		Ref.~\cite{Henriksson:2025vyi}.
		\\ \hline
		
		$\displaystyle \frac{g}{4!}\phi^{4}$
		&
		Satisfies the relation between anomalous dimension and beta-function
		in Eq.~\eqref{eq:check-phi4-beta}.
		\\ \hline
		
		$\displaystyle \frac{g^{2}}{5!}\phi^{5}$
		&
		Agreement through four loops with
		Ref.~\cite{Henriksson:2025vyi}, after accounting for the
		$2\beta(g)/g$ contribution from the operator normalization.
		\\ \hline
		
		Descendant and EOM operators
		&
		Consistent with the lower-dimensional renormalization
		relations derived in Section~\ref{re-descedant}.
		\\ \hline
		
	\end{tabular}
	\caption{Comparisons and consistency checks for the
		four-dimensional $\phi^{4}$ theory.}
	\label{tab:checks-phi4}
\end{table}

%%%%%%%%%%%%%%%%%%%%%%%%%%%%%%%%%%%%%%%%%%%
\paragraph{Internal consistency}
%%%%%%%%%%%%%%%%%%%%%%%%%%%%%%%%%%%%%%%%%%%
We perform several non-trivial internal consistency checks of the recursive
renormalization procedure. First, the renormalization of descendant and
equation-of-motion (EOM) operators is fixed by lower-dimensional operators.
For example, in the $\phi^{3}$ model, using
$E_{0}=Z_{\phi}^{-1/2}E$ and
$\partial^{2}\phi_{0}=Z_{\phi}^{1/2}\partial^{2}\phi$, the dimension-4
operator basis $(g_{0}\phi_{0}^{2}/2,\partial^{2}\phi_{0})$ gives
\begin{equation}
	\gamma_{\Delta=4}
	=
	\begin{pmatrix}
		-\gamma_{\phi} & -2\gamma_{\phi}\\
		0 & \gamma_{\phi}
	\end{pmatrix},
	\label{eq:check-descendant}
\end{equation}
which agrees with the corresponding entries obtained from the OPE
calculation. The same descendant and EOM relations reproduce the
corresponding entries in the higher-dimensional anomalous dimension
matrices.

Second, the tree-level OPE coefficient matrix has the block-triangular
structure shown in Table~\ref{block-triangular}. Each diagonal block has full
row rank. Therefore, the rank of the complete matrix is the number of hard
operators, and the recursive determination of the $Z$-factors is
non-degenerate at every step.

Third, the anomalous dimension of the marginal operator provides an
independent check involving the beta function. Denoting by
$\beta_{\mathrm{loop}}(g)$ the loop part of the beta function, the
dimension-6 $\phi^{3}$ result satisfies
\begin{equation}
	\gamma_{\phi^{3}}
	=
	g^{2}\frac{\partial}{\partial g}
	\left(\frac{\beta_{\mathrm{loop}}(g)}{g^{2}}\right)
	=
	-\frac{3g^{2}}{4}
	-\frac{125g^{4}}{48}
	+\mathcal{O}(g^{6}),
	\label{eq:check-phi3-beta}
\end{equation}
in agreement with the explicit OPE result. Similarly, for the marginal
operator in the $\phi^{4}$ model,
\begin{equation}
	\gamma_{\phi^{4}}
	=
	g\frac{\partial}{\partial g}
	\left(\frac{\beta_{\mathrm{loop}}(g)}{g}\right),
	\label{eq:check-phi4-beta}
\end{equation}
which is also satisfied by the anomalous dimension matrix obtained in this
work. These checks verify both the recursive closure of the operator basis
and the consistency of the extracted renormalization constants.
%Beyond comparisons with the literature, our framework provides several non-trivial internal consistency checks. (i) The renormalization of descendant and EOM operators is fully determined by lower-dimensional operators (Section~\ref{re-descedant}), and we have verified that all such entries in the anomalous dimension matrices agree with these relations. (ii) The upper-triangular block structure of the tree-level OPE coefficient matrix (Table~\ref{block-triangular}) ensures that diagonal blocks independently determine the corresponding $Z$-factors. (iii) In the $\phi^{3}$ model, the anomalous dimension of the marginal operator $\frac{g^{2}}{3!}\phi^{3}$ satisfies the beta-function relation~\eqref{phi3-dim6-beta}, which we have verified at two loops. 

%%%%%%%%%%%%%%%%%%%%%%%%%%%%%%%%%%%%%%%%%%%
\section{Conclusion}\label{conclusion}
%%%%%%%%%%%%%%%%%%%%%%%%%%%%%%%%%%%%%%%%%%%
In this work, we have extended the OPE-based renormalization algorithm to systematically handle operator mixing, a generic yet previously unaddressed challenge in the perturbative computation of anomalous dimensions. The central result is a recursive framework in which the $Z$-factors of arbitrary high-dimensional scalar ``hard'' operators $\Omega_{I}$ are fully and uniquely determined by the ultraviolet finiteness of OPE coefficients involving a strategically chosen basis of lower-dimensional symmetric traceless tensor ``soft'' operators $O_{\alpha}$.  Because the tree-level OPE coefficient matrix of this ``soft'' basis has full row rank, the $Z$-factors of higher-dimensional scalar operators can be derived iteratively from those of lower-dimensional ones, while the renormalization of intermediate tensor operators (such as conserved currents and descendants) is reduced to that of scalar operators through total-derivative combinations. Crucially, this ``global'' approach entirely eliminates the subtraction of UV or IR sub-divergences that burdens traditional methods such as the $R^{*}$  operation. As the first application of this framework, in the $\phi^{4}$ model, we have obtained the anomalous dimensions for scalar operators with $\Delta\leq 3$ to loop order five and for operators with $\Delta=4,5$ to loop order four. In the $\phi^{3}$ model, we obtain the two-loop anomalous dimensions for operators with $\Delta\leq 10$. 

These results establish the OPE-based method as a powerful and broadly applicable tool for multi-loop renormalization in the presence of operator mixing. We note that the current framework is limited to scalar hard operators in scalar field theories; extending it to fermionic operators and operators involving gauge fields will require further developments in the construction of the ``soft'' operator basis. A natural and important next step is the application to QCD, where operator mixing is ubiquitous and precision calculations are in high demand. Looking further ahead, the integration of this framework with conformal bootstrap techniques in CFT, its extension to higher-spin operators, and the push toward even higher loop orders all constitute promising directions for future research. In sum, the recursive OPE-based framework developed here transforms operator mixing from a computational bottleneck into a systematically solvable problem, opening a clear path toward precision calculations in QCD and beyond. 

%%%%%%%%%%%%%%%%%%%%%%%%%%%%%%%%%%%%%%%%%%%
\acknowledgments
%%%%%%%%%%%%%%%%%%%%%%%%%%%%%%%%%%%%%%%%%%%
We would like to thank Rijun Huang and Yi Li for inspiring discussions. This work is supported in part by the Science Challenge Project (No. TZ2025012), and NSAF No. U2330401.

%%%%%%%%%%%%%%%%%%%%%%%%%%%%%%%%%%%%%%%%%%%
\appendix
%%%%%%%%%%%%%%%%%%%%%%%%%%%%%%%%%%%%%%%%%%%
\section{The anomalous dimensions of marginal operators}\label{appendixA}
Let us take an example of $\phi^4$ model with single coupling. There are two marginal operators: 
\begin{equation}
    \begin{aligned}
        \Omega_{I}^{0}=\Bigl(\frac{1}{2}\partial_{\mu}\phi_{0}\partial^{\mu}\phi_{0},\ \frac{g_{0}}{4!}\phi_{0}^{4}\Bigr)\ .  
    \end{aligned}
\end{equation}
To find the anomalous dimensions of $\Omega_I$, we add them to the Lagrangian as perturbations 
\begin{equation}
    \begin{aligned}
        L_{\lambda}=L(\phi_{0},g_{0})+\lambda^{I}_{0}\Omega_{I}^{0}\ , 
    \end{aligned}
\end{equation}
and study the renormalization of the new theory in the $\mathcal{O}(\lambda)$ order. The parameters $\lambda_{0}^{I}$ can be written as 
\begin{equation}
    \begin{aligned}
        \lambda^{I}_{0}=\lambda^{J}{Z_{J}}^{I}(g)+\mathcal{O}(\lambda^{2})\ , 
    \end{aligned}
\end{equation}
in which $\lambda^{I}$ are renormalized parameters, and ${Z_{I}}^{J}$ are functions of $g$. 

Deriving correlation functions with respect to $\lambda^{I}$ yields finite quantities, 
\begin{equation}
    \begin{aligned}
        \frac{\partial}{\partial \lambda^{I}}\Bigl\langle \phi(x_{1})\cdots\phi(x_{n})\Bigr\rangle_{\lambda}=i\int d^{D}y\Bigl\langle {Z_{I}}^{J}\Omega_{J}^{0}(y)\phi(x_{1})\cdots\phi(x_{n})\Bigr\rangle\ . 
    \end{aligned}
\end{equation}
So ${Z_{I}}^{J}\Omega_{J}^{0}$ are renormalized operators: 
\begin{equation}
    \begin{aligned}
        \Omega_{I}^{R}={Z_{I}}^{J}\Omega_{J}^{0}\ . 
    \end{aligned}
\end{equation}
We can define new bare fields and coupling constants, so that the Lagrangian can be rewritten as 
\begin{equation}
    \begin{aligned}
        L_{\lambda}=L(\Phi_{0},\mathbf{g}_{0})\ . 
    \end{aligned}
\end{equation}
This still represents a $\phi^{4}$ model, but in terms of the new fields and coupling constants. In the new model, $Z$-factors still take the same form, but with $g$ replaced by $\mathbf{g}$: 
\begin{equation}\label{finite-quantity-1}
    \begin{aligned}
        \Phi_{0}=Z_{\phi}^{\frac{1}{2}}(\mathbf{g})\Phi,\ \ 
        \mathbf{g}_{0}=\tilde{Z}_{g}(\mathbf{g})\mathbf{g}\tilde{\mu}^{2\epsilon}\ . 
    \end{aligned}
\end{equation}
The bare and renormalized coupling constant $\mathbf{g}$ differ from the original coupling constants by $\mathcal{O}(\lambda)$ terms: 
\begin{equation}\label{g-expand}
    \begin{aligned}
        &\mathbf{g}_{0}=g_{0}+\delta g_{0}+\mathcal{O}(\lambda^{2})=g_{0}+c_{I}\lambda_{0}^{I}g_{0}+\mathcal{O}(\lambda^{2})\ ,\\
        &\mathbf{g}=g+\delta g+\mathcal{O}(\lambda^{2})\ , 
    \end{aligned}
\end{equation}
where $c_{I}$ are real numbers. Plug \eqref{g-expand} into \eqref{finite-quantity-1}, we find the following expression for $\delta g$: 
\begin{equation}\label{delta-g}
    \begin{aligned}
        &\frac{\delta g}{g}=\frac{1}{1+g\frac{\partial\ln \tilde{Z}_{g}}{\partial g}}\frac{\delta g_{0}}{g_{0}}=\Bigl( 1-\frac{\beta(g)}{2\epsilon g}\Bigr)\lambda
        ^{J}{Z_{J}}^{I}c_{I}\ . 
    \end{aligned}
\end{equation}
Since we are working in $\overline{MS}$ scheme, 
\begin{equation}
    \begin{aligned}
        {Z_{J}}^{I}=\delta_{J}^{I}+\text{terms with $\epsilon$-poles}\ , 
    \end{aligned}
\end{equation}
so the RHS of \eqref{delta-g} can be written as 
\begin{equation}
    \begin{aligned}
        \Bigl( 1-\frac{\beta(g)}{2\epsilon g}\Bigr)\lambda
        ^{J}{Z_{J}}^{I}c_{I}=\lambda^{I}c_{I}+\text{terms with $\epsilon$-poles}\ . 
    \end{aligned}
\end{equation}
But $\delta g/g$ is finite, so these terms with $\epsilon$-poles must vanish. We find
\begin{equation}\label{delta-g-3}
    \begin{aligned}
        &c_{I}\lambda^{I}=\Bigl( 1-\frac{\beta(g)}{2\epsilon g}\Bigr)\lambda
        ^{J}{Z_{J}}^{I}c_{I}\ . 
    \end{aligned}
\end{equation}
Since $\lambda^{I}$ are arbitrary parameters, \eqref{delta-g-3} infers 
\begin{equation}
    \begin{aligned}
    c_{I}=\Bigl( 1-\frac{\beta(g)}{2\epsilon g}\Bigr){Z_{I}}^{J}c_{J}\ . 
    \end{aligned}
\end{equation}
This means $c_{I}$ is an eigenvector of ${Z_{I}}^{J}$ with eigenvalue 
\begin{equation}
    \begin{aligned}
        z=\frac{1}{1-\frac{\beta(g)}{2\epsilon g}}\ . 
    \end{aligned}
\end{equation}
Consequently, it is also an eigenvector of ${\gamma_{I}}^{J}$: 
\begin{equation}
    \begin{aligned}
        {\gamma_{I}}^{J}c_{J}=-\frac{\partial \ln z}{\partial\ln\mu}c_{I}\ . 
    \end{aligned}
\end{equation}
So there must be an eigen-operator with anomalous dimension
\begin{equation}\label{gamma-beta}
    \begin{aligned}
        \gamma=-\frac{\partial \ln z}{\partial\ln\mu}=\frac{1}{1-\frac{\beta(g)}{2\epsilon g}}\frac{\partial g}{\partial \ln \mu}\frac{\partial }{\partial g}\Bigr(1-\frac{\beta(g)}{2\epsilon g}\Bigr)= g\frac{\partial }{\partial g}\frac{\beta(g)}{ g}\ . 
    \end{aligned}
\end{equation}
The eigen-operator is a linear combination of $g\phi^{4}$ and $(\partial\phi)^{2}$. 

Although we used $\phi^4$ model as an illustrative example, the major discussion here does not rely on the specific model. In a general model with a single coupling constant $g$, there exist an eigen-operator with anomalous dimension given by \eqref{gamma-beta}. For example, in 6-D $\phi^3$ model, the eigen-operator is a linear combination of $g\phi^3$ and $(\partial\phi)^{2}$, which gives the relation \eqref{phi3-dim6-beta}. In Yang–Mills theory, the eigen-operator is $\Tr(F^2)$, and the relation between its anomalous dimension and the beta function of the gauge coupling can be derived through a similar analysis~\cite{Grinstein:1988wz}. %In Yang-Mills theory, the eigen-operator is $\Tr(F^2)$, whose  relation is given in~\cite{Grinstein:1988wz}. 

\bibliographystyle{jhep}  % 使用 JHEP 规定的文献格式
\bibliography{ref2}  % 关联你的 .bib 文件（无需写扩展名）

% Bibliography

%% [A] Recommended: using JHEP.bst file
%% \bibliographystyle{JHEP}
%% \bibliography{biblio.bib}

%% or
%% [B] Manual formatting (see below)
%% (i) We suggest to always provide author, title and journal data or doi:
%% in short all the informations that clearly identify a document.
%% (ii) please avoid comments such as "For a review'', "For some examples",
%% "and references therein" or move them in the text. In general, please leave only references in the bibliography and move all
%% accessory text in footnotes.
%% (iii) Also, please have only one work for each \bibitem.

%\begin{thebibliography}{99}

%\bibitem{a}
%Author,
%\emph{Title},
%\emph{J. Abbrev.} {\bf vol} (year) pg.

%\bibitem{b}
%Author,
%\emph{Title},
%arxiv:1234.5678.

%\bibitem{c}
%Author,
%\emph{Title},
%Publisher (year).

%\end{thebibliography}
\end{document}